\UseRawInputEncoding
\documentclass[preprint,12pt]{elsarticle}

\usepackage{amssymb}
\usepackage{amsmath}

\usepackage{booktabs}
\usepackage{hyperref}  
\usepackage{multirow}
\usepackage{rotating}

\usepackage{pdflscape} 
\usepackage{adjustbox} 

\usepackage{changes}

\journal{Journal of Systems and Software}

\begin{document}

\begin{frontmatter}



\title{Domain-Driven Design in Software Development: A Systematic Literature Review on Implementation, Challenges, and Effectiveness}


\author[TUe]{Ozan \"{O}zkan}
\author[TUe,WUR]{\"{O}nder Babur}
\author[TUe]{Mark van den Brand}

\affiliation[TUe]{organization={Mathematics and Computer Science, Eindhoven University of Technology},
            city={Eindhoven},
            country={The Netherlands}}
\affiliation[WUR]{organization={Information Technology Group, Wageningen University and Research},
            city={Wageningen},
            country={The Netherlands}}

\begin{abstract}
\textbf{Context:} Domain-Driven Design (DDD) has gained significant attention in software development for its potential to address complex software challenges, particularly in the areas of system refactoring, reimplementation, and adoption. Using domain knowledge, DDD aims to solve complex business problems effectively. \\
\textbf{Objective:} This SLR aims to provide an analysis of existing research on DDD in software development, paint a picture of DDD in solving software problems, identify the challenges encountered during its application and explore the results of these studies. \\
\textbf{Method:} We systematically selected 36 peer reviewed studies and conducted quantitative and qualitative analyzes to synthesize the findings. \\
\textbf{Results:} DDD has effectively improved software systems, with its key concepts. The application of DDD in microservices has gained prominence for its ability to facilitate system decomposition. Some studies lacked empirical evaluations, highlighting challenges in onboarding and the need for expertise. \\ 
\textbf{Conclusion:} Adopting DDD benefits software development, involving stakeholders such as engineers, architects, managers, and domain experts. More empirical evaluations and open discussions on challenges are needed. Collaboration between academia and industry advances the adoption and transfer of knowledge of DDD in projects.
\end{abstract}

\begin{graphicalabstract}
\end{graphicalabstract}

\begin{highlights}
\item This Systematic Literature Review (SLR) on Domain-Driven Design (DDD) highlights its potential benefits in software development and the importance of involving key stakeholders for successful implementation.
\item Studies suggest a growing interest and adoption of DDD in the context of microservices since 2017. However, more empirical research is needed to fully understand the benefits and challenges of implementing DDD in different software development scenarios.
\item The study reveals variations in the implementation and evaluation of the DDD principles in the included studies. 
\item To improve the quality of DDD research, future studies should focus on the consistent use of DDD principles, robust evaluation methodologies, and open discussion of both advantages and limitations. This approach can provide a comprehensive understanding of DDD's potential and practical implications
\item The implementation of DDD significantly depends on the expertise of stakeholders. Experienced developers and domain experts are crucial to effectively applying DDD concepts and practices.
\item Collaboration between academia and industry plays a crucial role in advancing the knowledge and adoption of DDD. Academic studies contribute to theoretical advances, while industry-focused studies showcase practical implementations and real-world applicability.
\end{highlights}


\end{frontmatter}



\section{Introduction}

In the field of software development, the pursuit of effective design methodologies and architectural approaches has been a crucial endeavor for both researchers and practitioners. Among these approaches, DDD has gained significant attention and adoption, particularly in the context of solving software architecture problems and addressing challenges related to refactoring, reimplementation and adoption \cite{avram_domain-driven_2006}. DDD, as advocated by Eric Evans in his influential book, \textit{"Domain-Driven Design: Tackling Complexity in the Heart of Software"} \cite{evans_domain-driven_2004}, offers a framework for understanding, modeling, and solving complex business problems by placing domain knowledge at the core of the design process.

DDD emphasizes the alignment of software systems with the underlying business domain, enabling developers to better understand and model complex business requirements. By capturing the language, concepts, and relationships within the domain, DDD helps to identify and address software architecture challenges effectively \cite{evans_domain-driven_2004, vernon_implementing_2013}. This approach provides guidance on how to improve system quality, maintainability, and scalability by ensuring that software design closely mirrors the business domain \cite{evans_domain-driven_2004}.

Although DDD presents solutions with its conceptual elegance and potential benefits in managing software architecture challenges \cite{evans_domain-driven_2004, vernon_implementing_2013}, it is crucial to evaluate its practical application and assess its impact on software development projects. This Systematic Literature Review (SLR) aims to provide a comprehensive analysis of existing research studies that have used DDD for various purposes in software development, such as improving system design, enhancing maintainability, and facilitating communication between stakeholders. By analyzing and synthesizing empirical studies, we aim to gain insight into the effectiveness of DDD in solving software architecture problems and identify the challenges encountered during the application of DDD from these studies.

To achieve this objective, a rigorous research methodology was employed that adheres to established guidelines proposed by Kitchenham et al. \cite{kitchenham_guidelines_2007} and Wohlin et al. \cite{wohlin_guidelines_2014}. A group of 36 peer-reviewed studies was systematically selected and quantitative and qualitative analyzes were performed based on the data extracted from these studies.

The conclusion drawn from this SLR emphasizes the potential benefits of adopting DDD in software development. The study highlights the importance of involving key stakeholders, including software engineers, architects, project managers, and domain experts, to ensure a successful implementation. In addition, the study identifies the importance of open discussions on challenges and limitations and emphasizes the collaboration between academia and industry to advance the adoption of DDD and the transfer of knowledge in real-world projects.

This study contributes to the growing body of knowledge on DDD and provides valuable insights for practitioners and researchers seeking to apply the principles of DDD effectively in software development. By identifying the strengths and weaknesses of existing research and proposing areas for improvement, this SLR aims to foster the further advancement and adoption of DDD principles in contemporary software development projects.

\subsection{Domain-Driven Design Essentials} \label{subsec:domain_driven_design}
DDD is a conceptual framework and approach to software development in software engineering that seeks harmony between the structure of software code and the complexities of the business domain it serves. The goal of DDD is to bridge the gap between technical implementation and business requirements, resulting in more effective communication between developers and domain experts \cite{evans_domain-driven_2004, vernon_implementing_2013}. This concept was pioneered by Eric Evans, whose significant book \textit{Domain-Driven Design: Tackling Complexity in the Heart of Software} \cite{evans_domain-driven_2004} has shaped this paradigm. Evans emphasizes the alignment of software design with the intricate nuances of the business domain. He argues that effective software development requires a deep understanding of the complexities of the domain, allowing developers to create software that genuinely resonates with real-world needs \cite{evans_domain-driven_2004}.  

In DDD, the domain refers to the specific area of expertise or the subject matter of the software application. This could be anything from e-Commerce, finance, healthcare, or logistics. The central concept in DDD is the development of a rich, well-defined domain model that accurately represents the business rules, processes, and entities within the domain. This model is built through collaboration between developers and domain experts, often using a shared language called \textit{Ubiquitous Language} that captures both technical and domain-specific terms.

Through its concepts, which are explained in the following sections, DDD introduces principles that help software architects effectively represent the complex aspects of the real world. By embracing these core tenets, software engineers navigate the complex landscape of software development with a compass calibrated to the nuances of the domain, thus achieving a symbiotic fusion of technical precision and domain understanding.

\subsubsection{Ubiquitous Language}
At the core of DDD lies the concept of Ubiquitous Language, a practice that aims to bridge the gap between technical jargon and domain-specific vocabulary \cite{evans_domain-driven_2004}. Ubiquitous Language fosters a shared vocabulary that harmonizes conversations among developers, domain experts, and business stakeholders. By encapsulating the complexities of the business domain in a common understanding, Ubiquitous Language eliminates ambiguity and streamlines communication. This shared linguistic framework empowers stakeholders to articulate domain intricacies effortlessly and translates directly into the software's code, facilitating a direct and unambiguous translation of domain understanding into the software's design and implementation.

\subsubsection{Bounded Context}
The Bounded Context concept effectively sets the boundaries of distinct zones where the semantics of a domain model are consistently applied. In applications of substantial scale that incorporate a variety of domain models, there is an inherent risk of ambiguity and misalignment. Bounded Contexts serve as strategic divisions, separating different domain models to maintain their integrity. This approach confines each domain to specific contexts, thus preventing semantic conflicts and ensuring a unified representation of the business domain within each delimitated area \cite{evans_domain-driven_2004}. It can serve as an abstraction of a (sub-)system or team, aligning boundaries with software components, organizational structures, and responsibilities.

\subsection{Context Mapping} \label{subsec:context_mapping}
In DDD, Context Mapping is a strategic practice used to understand and manage the relationships between different bounded contexts in complex software systems \cite{evans_domain-driven_2004}. A Bounded Context defines the boundary within which a particular domain model is defined and applicable. However, in real-world systems, multiple teams often work on different subdomains or systems, each with its own Bounded Context. These contexts must interact and integrate, which introduces the risk of model contamination or ambiguity across teams.

Context Mapping provides a visual and conceptual technique to explicitly define how bounded contexts relate to each other and how integration should be handled. It helps development teams determine appropriate integration patterns (e.g., Shared Kernel, Customer/Supplier, Conformist, or Anti-Corruption Layer) depending on the nature of team collaboration, legacy constraints, or organizational structure. These relationships are represented using a Context Map, which becomes an essential communication artifact in large systems involving multiple teams and domains.

Evans emphasizes that strategic design decisions are essential when teams develop models in parallel, and that context mapping serves as a bridge to align models and avoid conflicts \cite{evans_domain-driven_2004}. In practice, context maps also guide technical implementation, by informing decisions about API design, data ownership, and service integration. As such, context mapping is a vital part of ensuring semantic consistency and architectural integrity across distributed systems.

\subsubsection{Aggregates}
The concept of Aggregates encapsulates Entities, Value Objects, and Domain Logic through methods in entities and the use of domain services. Aggregates are the main points for accessing and changing data, ensuring data integrity within business transactions \cite{evans_domain-driven_2004}. By containing data within defined boundaries, Aggregates optimize data management and system scalability while also enforcing logical boundaries that prevent uncontrolled data manipulation.

\subsubsection{Entities and Value Objects}
Entities and Value Objects differentiate within the DDD framework. Entities embody tangible elements within the domain, each possessing a unique identity. These entities change over time, capturing the mutable aspects of the business reality \cite{evans_domain-driven_2004}. In contrast, Value Objects represent conceptual attributes that lack identity and remain immutable. By distinguishing between these two constructs, DDD provides a complete representation of the domain, accurately modeling tangible entities and abstract attributes \cite{evans_domain-driven_2004}.

\subsubsection{Domain Services}
Domain Services encapsulate domain logic that transcends individual entities and value objects \cite{evans_domain-driven_2004}. These services orchestrate complex operations and interactions within the domain. By segregating multifaceted logic, domain services promote modular, reusable, and concise code. This approach mirrors the coherent coordination of real-world processes, enhancing the software's alignment with the domain's intricacies.

\subsubsection{Domain Events}
Domain Events is a concept used to model and communicate changes or significant occurrences within a domain \cite{vernon_implementing_2013}. They play a role in capturing and representing the behavior and state changes within an application's core domain. A Domain Event is a lightweight object that represents something important that has happened in the domain. These events are usually named in the past tense to indicate that they've already occurred. They help decouple different parts of a system by allowing different components or aggregates to communicate without needing to know the specifics of each other. This promotes a more modular and maintainable architecture.
 
\subsubsection{Anti-Corruption Layer (ACL)}
The concept of an ACL protects the domain model from external influences. This layer mediates interactions between different models or systems, preserving the domain's integrity. By translating data between disparate representations, the ACL ensures a consistent domain model, safeguarding it from external distortions \cite{evans_domain-driven_2004}.

\subsubsection{Core Domain}
Core Domain is a central concept of strategic design in DDD, which refers to the part of the software system that delivers the most value and differentiates the business. The goal is to focus development effort on the Core Domain to gain competitive advantage, often assigning the most skilled developers to it \cite{evans_domain-driven_2004}.

\subsubsection{Shared Kernel}
Shared Kernel refers to a pattern in which two Bounded Contexts share a common subset of the domain model. This shared part must be carefully managed to avoid conflicts and preserve the integrity of the model, often requiring close collaboration and coordination between teams \cite{evans_domain-driven_2004}.

The ACL shares similarities with several well-known design patterns. For instance, the \textit{Remote Proxy} pattern \cite{gamma_design_2009} provides a local representative for an object that resides in a different address space, mediating interactions between local and remote systems. The \textit{Wrapper} pattern \cite{buschmann_pattern-oriented_1996} encapsulates an object to provide a different interface, much like the ACL encapsulates external systems to present a consistent interface to the domain model. The \textit{Adapter} pattern \cite{gamma_design_2009} converts the interface of a class into another interface that a client expects, similar to how the ACL adapts external data representations to match the internal domain model.

The fundamental contribution of DDD lies in creating software systems that closely mirror the business's cognitive schema. By converging domain experts, business stakeholders, and software developers, DDD facilitates synergistic collaboration to deliver software solutions that optimally cater to business needs. This collaborative effort fosters the combination of domain insights, augments the comprehension of problems and solutions across the entirety of the development team, and supports the fostering of inter-team relationships.

\subsubsection{Strategic and Tactical DDD}
Over time, the DDD community has categorized DDD patterns into two broad groups: \textit{Strategic} and \textit{Tactical}. Evans also has a similar cateogization in his book \cite{evans_domain-driven_2004} as he calls it \textit{Strategic Design}, but the community calls it as \textit{Tactical Design} as of today. It has become a widely adopted framing, especially in practitioner literature such as Vernon’s work \cite{vernon_implementing_2013}. Strategic DDD focuses on aligning the software model with the broader business strategy and organization, with key concepts including Bounded Context, Core Domain, Context Maps, and Anti-Corruption Layer. In contrast, Tactical DDD includes concrete modeling techniques used within a Bounded Context, such as Entities, Value Objects, Aggregates, Domain Services, and Domain Events. Throughout this study, we use this distinction as a lens to interpret the usage patterns of DDD concepts in the literature.

\subsubsection{Event Storming}
In addition to modeling patterns, DDD also includes collaborative modeling techniques that help bridge the gap between business and technical stakeholders. One widely adopted practice is Event Storming, a workshop-based approach introduced by Alberto Brandolini \cite{brandolini_eventstorming_2013}, where domain experts and developers collaboratively explore business processes by identifying and organizing domain events, commands, aggregates, and actors. This method is particularly effective for discovering Bounded Contexts and aligning the domain model with real-world business operations.

\subsubsection{Other DDD Patterns and Concepts}
The DDD patterns and concepts introduced in this section reflect those discussed most frequently in the 36 primary studies reviewed. Although there are additional patterns in the DDD literature, such as Customer/Supplier, Open Host Service, Conformist, and Published Language, these were not consistently represented in the reviewed studies and, therefore, were not included in detail. Our goal in this section is to provide a focused overview of the key patterns that emerged from the literature rather than an exhaustive catalog of all DDD concepts.

\section{Related Work} \label{sec:related_work}
To the best of our knowledge, despite the growing interest in DDD as a software development approach, there is a scarcity of SLRs specifically focused on DDD. While the literature landscape on DDD is not completely unexplored, previous studies have primarily touched upon certain aspects remotely related to DDD. 

For instance, Le et al. \cite{le_domain-driven_2016} propose a set of concrete design patterns for Domain Modeling within the context of Object-Oriented DDD, providing valuable insights into the practical application of DDD in specific modeling scenarios. Singjai et al.~\cite{singjai_practitioner_2021}, on the other hand, examine the interrelation between microservice APIs and DDD, shedding light on how DDD principles can be applied in the context of microservices architecture. However, it is worth noting that some related systematic literature reviews have been conducted in adjacent domains. Goey et al.~\cite{de_goey_design-driven_2019}, for instance, undertakes a Systematic Literature Review on Design-Driven Innovation, exploring design-driven approaches and methodologies in innovation processes, which, although related to design-driven practices, do not directly address the specific principles and patterns of DDD. Similarly, Bertoni et al.~\cite{bertoni_data-driven_2020} conducts an SLR on Data-Driven Design in Concept Development, focusing on the role of data-driven approaches in the early stages of concept development but not explicitly addressing the nuances of DDD. 

Despite valuable contributions from these studies, none of them provide a comprehensive review of DDD as a whole, addressing its principles, methodologies, and practical applications. While these prior studies offer valuable perspectives related to design-driven or microservice-oriented software development, they do not conduct a structured or comprehensive SLR focused on the conceptual and practical aspects of DDD itself. In contrast, our study takes a DDD-centric perspective, systematically analyzing how DDD principles, patterns, and stakeholders are discussed and applied in peer-reviewed literature. This focus enables us to capture insights specifically tied to DDD adoption, effectiveness, and evaluation, filling a gap not directly addressed by adjacent literature.

\section{Research Objectives and Method} \label{sec:research_objectives_and_method}
This study employed a Systematic Literature Review methodology to create generalizable empirical results (as opposed to specific and in-depth yes less generalizable research methods such as case studies~\cite{stol2018abc}) to (1) consolidate the existing knowledge on the utilization of DDD in the domain of software development, (2) identify gaps in the current research, guiding future studies and helping to focus on underexplored areas, and (3) for practitioners, a comprehensive review can offer insights into best practices, common challenges, and effective strategies for implementing DDD in real-world projects. The research approach adhered to established guidelines as proposed by Kitchenham et al. \cite{lubke_interface_2019} and Wohlin et al. \cite{wohlin_guidelines_2014}.

Additionally, this study delves into further detail regarding the Goal and Research Questions (Section \ref{subsec:goal_and_research_questions}), Search Strategy and Data Sources (Section \ref{subsec:search_strategy_and_data_sources}), Exclusion Criteria (Section \ref{subsec:exclusion_criteria}), Primary Study Identification and Selection (Section \ref{subsec:primary_study_identification_and_selection}), and Data Extraction (Section \ref{subsec:data_extraction}) in subsequent sections. By examining these aspects in a meticulous manner, this study aims to provide a comprehensive and rigorous analysis of the research topic at hand.

\subsection{Goal and Research Questions} \label{subsec:goal_and_research_questions}
The primary objective of this research is to investigate and synthesize the current state of research on the application of DDD in software development, to understand its benefits, challenges, used patterns, and effectiveness. To achieve this objective, the following research questions (RQ)s have been formulated: 

\begin{itemize}
    \item []\textbf{RQ1:} What is the state of the art concerning DDD in the existing literature, in terms of study types, research venues, software systems, and the distribution between academic and industrial settings?
    \item[]\textbf{RQ2:} What categories of software systems have demonstrated benefits from the implementation of DDD?
    \item[]\textbf{RQ3:} Which specific software development concerns have been addressed through the application of DDD in previous studies?
    \item[]\textbf{RQ4:} What are the most common DDD patterns utilized in previous research, such as ACLs, Bounded Contexts, and other related approaches?
    \item[]\textbf{RQ5:} What are the challenges encountered during the implementation of DDD in software development projects?
    \item[]\textbf{RQ6:} How has the effectiveness of DDD been assessed and measured in prior studies?
    \item[]\textbf{RQ7:} Who are the key stakeholders involved and benefited in the implementation of DDD?
\end{itemize}

\subsection{Search Strategy and Data Sources} \label{subsec:search_strategy_and_data_sources}
In line with established protocols and our previous studies \cite{giray_use_2023,ochoa2025characterising}, we meticulously designed our search strategy to identify pertinent studies from reputable academic databases renowned for their extensive coverage of computer science and software engineering literature. The databases utilized in this study were \textit{ACM}, \textit{SpringerLink}, \textit{IEEE}, and \textit{Wiley}, all of which are widely recognized for their authoritative and diverse repositories of articles.

The search query employed in this investigation was thoughtfully crafted to encompass a broad spectrum of articles relevant to our research objectives. The query utilized a combination of keywords and logical operators to target studies explicitly discussing DDD in conjunction with software development activities. We used "Domain-Driven Design" because we explicitly wanted to capture studies that employed DDD. When using only "Domain-Driven Design", we encountered a significant number of false positives, such as studies from medicine. Therefore, to ensure the relevance of the captured studies, we included implementation keywords to specifically address studies that employed DDD in practice. The search query was formulated as follows:

\begin{center}
\begin{verbatim}
"Domain Driven Design" AND (refactor* OR improv* OR optimiz* 
OR restructur* OR migrat* OR evolution OR cleanup OR 
reengineering)
\end{verbatim}
\end{center}

We strategically used asterisks (*) as wildcards to capture variations of the specified keywords and to ensure inclusivity in the search results.

\subsection{Exclusion Criteria} \label{subsec:exclusion_criteria}
The establishment of selection criteria was undertaken with the aim of mitigating bias and minimizing subjectivity within the study. The inclusion and exclusion criteria were formulated in accordance with the research questions (RQs) as delineated by Kuhrmann et al. \cite{kuhrmann_pragmatic_2017}, thereby ensuring that the criteria align closely with the intended objectives of the investigation. Following our criteria, (1) we meticulously identified and removed duplicate studies that appeared in more than one of the selected databases to prevent the inclusion of redundant information. To maintain the comprehensiveness of our analysis, (2) we excluded studies that lacked full-text availability, as the absence of complete content may hinder a thorough evaluation of the study's findings and methodology. Given the language limitations of our review team, (3) we restricted our analysis to studies written in English to ensure accurate and consistent interpretation and analysis. In the interest of focusing on primary research articles, (4) we excluded books, book chapters, editorials, and issue introductions from our analysis, as they may not present original empirical studies or detailed investigations. (5) We excluded studies that were not primary research studies (e.g., review articles, meta-analyses, opinion pieces) to maintain a clear focus on empirical research directly relevant to our research questions. To ensure the inclusion of studies closely related to DDD and its practical applications, (6) we excluded studies that did not explicitly present or discuss the introduction of a method, technique, tool, or evaluation pertaining to DDD or its adaptation and extension. The list of the exclusion criteria can be seen in Table \ref{tab:exclusion_criteria}.

\begin{figure}[t]
  \centering
  \includegraphics[width=\linewidth]{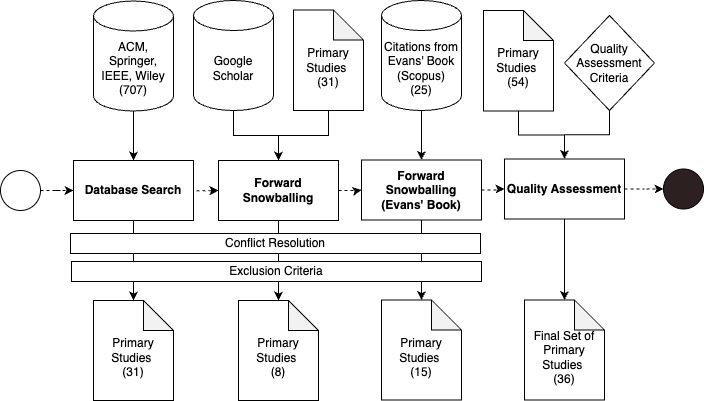}
  \caption{Study selection progressed from a database search to the quality assessment. The final set of primary studies advanced to the data extraction stage. The count of studies in the evaluation steps is displayed in parentheses.}
  \label{fig:study_selection_process}
\end{figure}

\begin{table}[t]
  \centering
  \begin{tabular}{cl}
    \hline
    \ & Exclusion Criteria \\
    \hline
    EC1 & Duplicate papers from multiple sources. \\
    EC2 & Papers without full text available. \\
    EC3 & Papers not written in English. \\
    EC4 & Books, book chapters, editorials, issue introductions. \\
    EC5 & Papers are not primary studies. \\
    EC6 & Papers do not explicitly introduce a method, technique, \\
    & tool and/or evaluation of adaptation or extension. \\
    \hline
    \hline
    \ & Inclusion Criterion \\
    \hline
     & Papers do not meet any exclusion criteria. \\
    \hline
  \end{tabular}
  \caption{List of exclusion and inclusion criteria.}\label{tab:exclusion_criteria}
\end{table}

\subsection{Primary Study Identification and Selection} \label{subsec:primary_study_identification_and_selection}
\subsubsection{Conducting Manual Database Search} \label{subsubsec:conducting_manual_database_search}
In the first stage, we performed a meticulous and systematic manual search of the databases mentioned in Section 2.2. The search was carried out in March 2023 and yielded a total of 707 studies, which were 153 studies from ACM, 19 studies from IEEE Xplore, 510 studies from SpringerLink, and 25 studies from Wiley, as also shown in Table \ref{tab:primary_study_identification_metrics}. To ensure the exclusion of duplicate results, we conducted conflict resolution, resulting in the identification of one duplicate paper. After eliminating duplicates, we advanced 706 unique studies to the next stage of the selection process.

\subsubsection{Applying Exclusion Criteria} \label{subsubsection:applying_exclusion_criteria}
In the second stage, we applied our predefined exclusion criteria to the 706 studies. The screening process involved the assessment of each paper based on its title, abstract, and keywords. In situations where a thorough evaluation was not possible based on these elements, we briefly reviewed the introduction and conclusion of the paper, following the guidelines suggested by Wohlin et al. \cite{wohlin_guidelines_2014}. To ensure a rigorous assessment, we adopted a multi-assessor strategy, where 2 authors independently evaluated the studies without knowing each other's assessments. Due to the high number of studies, we performed the assessment in iterations until we achieved at least 0.80 Cohen's Kappa coefficient \cite{cohen_coefficient_1960}, which indicates almost perfect agreement between the assessors \cite{cohen_coefficient_1960}. Once the desired Cohen's Kappa was reached, the first author continued to assess the remaining studies to determine their inclusion or exclusion, following the approach done in previous studies \cite{torres_systematic_2021, wang_machinedeep_2023}. 
In each iteration, we randomly selected 30 studies for assessment. After each iteration, we conducted conflict resolution sessions to discuss the rationale of our assessments. We eventually achieved 1.0 Cohen's Kappa coefficient in the second iteration. The remaining studies were evaluated by the first author, resulting in 31 studies marked for inclusion and progression to the next stage. The distribution of these included studies across the search databases is presented in Table \ref{tab:primary_study_identification_metrics}.

Table \ref{tab:exclusion_breakdown} provides a detailed breakdown of the number of studies excluded under each exclusion criterion. This complements the overall metrics shown in Table~\ref{tab:primary_study_identification_metrics}, providing transparency on how rigorously the exclusion process was applied.

\begin{table}[t]
  \centering
  \begin{tabular}{cl}
    \hline
    \ & Count of studies excluded per criterion \\
    \hline
    EC1 & 11 studies excluded as duplicates across databases. \\
    EC2 & 15 studies excluded due to lack of full-text access. \\
    EC3 & 0 studies excluded for being in non-English languages. \\
    EC4 & 461 studies excluded as books, editorials, or chapters. \\
    EC5 & 53 studies excluded as they were not primary studies. \\
    EC6 & 147 studies excluded for not introducing or evaluating \\ 
    & DDD methods, tools, or adaptations. \\
    \hline
    Total & 687 studies excluded across all criteria. \\
    \hline
  \end{tabular}
  \caption{Number of studies excluded per criterion during the screening process.}
  \label{tab:exclusion_breakdown}
\end{table}

\subsubsection{Conducting Forward Snowballing} \label{subsubsec:conducting_forward_snowballing}
In the third stage, we conducted forward snowballing to make our study population wider and ensure relevancy. We conducted the forward snowballing in two approaches. We conducted a forward snowballing process for studies that cited Evans' book \cite{evans_domain-driven_2004} since Evans holds significant prominence in the field of DDD due to his influential contributions. His book, which serves as a reference, has gained substantial recognition and is widely regarded as a crucial resource among practitioners in the software development community \cite{avram_domain-driven_2006, vernon_implementing_2013, vernon_domain-driven_2016, nilsson_applying_2006}. To conduct this snowballing, we utilized Scopus as a database because of its centralized indexing of our target databases mentioned in Section \ref{subsec:search_strategy_and_data_sources} and to eliminate results from grey literature. The screening process for Evans' book forward snowballing followed the same strategy as that used for forward snowballing in the primary studies. In the process of analysis, a comparison was made among the database search results, forward snowballing results, and the previously identified studies to identify any potential duplicates. As a result, a total of 10 duplicate studies were identified, and these duplicates were subsequently resolved in favor of the database search results and forward snowballing outcomes. Consequently, an additional 15 studies that did not meet any of the exclusion criteria were identified and subsequently included in our study. The comprehensive forward snowballing process, encompassing both the forward snowballing of included studies and the examination of studies cited in Evans' book, resulted in a total of 23 studies being incorporated into our analysis. 

We also performed forward snowballing following Wohlin et al. guidelines \cite{wohlin_guidelines_2014}, we used Google Scholar to identify studies that cited our included primary studies. We assessed these additional studies based on their title, abstract, and keywords, using the exclusion criteria established earlier. In cases where a definitive resolution was not achievable based on the provided information, we conducted a rapid analysis of the entire paper, as suggested by Wohlin et al. \cite{wohlin_guidelines_2014}. Through this process, we identified 8 studies that did not meet any of the exclusion criteria and were therefore included in our study.

\subsubsection{Quality Assessment} \label{subsubsec:quality_assessment}
During stage 4 of our primary study identification and selection process, we proceeded with the task of conducting a comprehensive quality assessment of the selected studies. Adhering to the established guidelines put forth by Kitchenham et al. \cite{kitchenham_guidelines_2007}, each study was meticulously evaluated, with ratings assigned on a scale ranging from 0 to 1. On this scale, a rating of 1 denoted a positive affirmation, 0.5 indicated a partial fulfillment of the quality criteria, and 0 represented a negative response. The specific quality assessment questions utilized in this evaluation can be found in Table \ref{tab:quality_criteria}. Studies that achieved a total score of 4 or higher were deemed to have successfully met the quality assessment criteria and were subsequently included in the data extraction stage. In order to maintain the rigor and objectivity of the process, we adopted a multi-assessor strategy during the quality assessment stage. Each study was independently evaluated by two assessors. In cases of disagreement or doubt, the assessors consulted with each other to reach a consensus. This approach ensured that the quality assessments were rigorous, minimized potential biases, and allowed discrepancies to be resolved systematically. After quality assessment, we identified the 36 primary studies. The final list of primary studies can be found in Table \ref{tab:list_of_primary_studies}.

Among the 36 selected studies, the majority demonstrated strong adherence to methodological rigor: 78\% received a full score (1.0) for clearly stating research aims (QA1), and 75\% scored full on explaining their methodology (QA2). However, some common weaknesses were observed. For instance, 39\% of studies scored partially (0.5) or zero on reporting negative findings (QA5), indicating a tendency to report only positive outcomes. Additionally, 36\% had incomplete descriptions of their data collection or analysis procedures (QA3), which limits reproducibility. These gaps reflect typical challenges in empirical DDD research, where reporting standards may vary, and certain studies lean more toward industrial experience than academic evaluation. Despite this, all selected studies met a minimum quality threshold, ensuring a consistent baseline for analysis.

\begin{table}[t]
  \centering
  \begin{tabular}{cl}
    \hline
    \# & Quality Criteria \\
    \hline
        QC1 & The aim of the study is clearly stated. \\
        QC2 & The scope and context of the study are clearly defined. \\
        QC3 & The study is variable, valid and reliable. \\
        QC4 & The research process of the study is documented properly. \\
        QC5 & The research questions in the study are answered. \\
        QC6 & Negative findings in the study are presented. \\
        QC7 & The main findings are clearly stated. \\
        QC8 & Conclusions relate to the aim and purpose of the study. \\
    \hline
  \end{tabular}
  \caption{List of quality criteria adopted from Kitchenham et al. guidelines \cite{kitchenham_guidelines_2007}. Each criterion is rated on a scale of 0 to 1. Studies that achieved a score of $>=$4.0 advanced to the data extraction stage.}
  \label{tab:quality_criteria}
\end{table}

\begin{table}[t]
  \centering
  \begin{tabular}{lccccc}
    \hline
    Database & Search Date & Results & Conflict & Included & Quality C. \\
    \hline
    ACM & March 2023 & 153 & 0 & 13 & 12 \\
    IEEE Xplore & March 2023 & 19 & 0 & 8 & 3 \\
    SpringerLink & March 2023 & 510 & 1 & 8 & 5 \\
    Wiley & March 2023 & 25 & 0 & 2 & 1 \\
    \hline
    Sum Database & & 707 & 1 & 31 & 21 \\
    \hline
    Scopus & April 2023 & 608 & 10 & 15 & 9 \\
    Google Scholar & June 2023 & 93 & 0 & 8 & 6 \\
    \hline
    Sum Snowball & & 701 & 10 & 23 & 15 \\
    \hline
    Total & & 1408 & 11 & 54 & 36 \\
    \hline
  \end{tabular}
  \caption{Primary study identification metrics showing accessed databases, search dates, number of resulting studies, conflicted studies, included studies, and the count of studies that passed quality checks.}
  \label{tab:primary_study_identification_metrics}
\end{table}

\subsection{Data Extraction and Synthesis} \label{subsec:data_extraction}
\subsubsection{Data Extraction}
In order to ensure consistency and alignment with our research questions and goals, a data extraction form was developed (Table \ref{tab:data_extraction_form}). The initial list of data points was prepared by the first author, and this list was then discussed among the authors to refine and finalize a starting set of categories. A pilot data extraction was done on a subset of randomly selected primary studies. During the data extraction process, we adopted an iterative and incremental approach, continuously refining the categories as new insights emerged. This iterative refinement allowed us to adapt the data extraction form to better capture the nuances and specific details relevant to our RQs and goals.

The data points were categorized into 3 main categories. Firstly, the \textit{Publication Metadata} category involved extracting basic information such as the title, authors, and publisher of each study. Secondly, the \textit{Study Characteristics} category included data related to the study design, data collection methods, and evaluation metrics employed. Lastly, the \textit{Domain-Driven Design} category focused on extracting information specifically pertaining to the application of DDD in each study.

Following the completion of the quality assessment stage, the final set of studies deemed relevant and has sufficient quality to be moved to the data extraction stage. The data extraction process was carried out by the first author using a predefined data extraction form. This form served as a structured tool to systematically collect and record pertinent information from the selected studies.

\begin{table}[htbp]
\centering
\begin{tabular}{clcc}
  \hline
  Category & Field & Input Type & RQ \\
  \hline
    \multirow{8}{*}{\rotatebox[origin=c]{90}{\textit{Publication Metadata}}}
    & ID & Auto Alphanumeric & - \\
    & Title & Free Text & RQ1 \\
    & Author & Free Text & RQ1 \\
    & Item Type & Selection & RQ1 \\
    & Database & Selection & RQ1 \\
    & Publication Year & Number & RQ1 \\
    & Publisher & Free Text & RQ1 \\
    & Source & Selection & RQ1 \\
    \\
    \multirow{9}{*}{\rotatebox[origin=c]{90}{\textit{Study Characteristics}}}
    & Study Design & Selection & RQ1 \\
    & Data Collection Methods & Free Text & RQ1 \\
    & Academic/Industry & Binary & RQ1 \\
    & Participants & Free Text & RQ7 \\
    & Empirically Evaluated & Binary & RQ1 \\
    & Human Metrics & Binary & RQ1 \\
    & Evaluation Methods & Free Text & RQ6 \\
    & Evaluation Metrics & Free Text & RQ6 \\
    & Evaluation Results & Free Text & RQ6 \\
    \\
    \multirow{5}{*}{\rotatebox[origin=c]{90}{\textit{DDD (Context)}}}
    & Software System & Free Text & RQ2 \\
    & Problems/Motivations & Free Text & RQ3 \\
    & Evaluation of Effectiveness & Free Text & RQ6 \\
    & Challenges and Limitations & Free Text & RQ5 \\
    & Lessons Learned & Free Text & RQ4-5 \\
    \\
    \multirow{7}{*}{\rotatebox[origin=c]{90}{\textit{DDD (Content)}}}
    & Ubiquitous Language & Binary & RQ4 \\
    & Bounded Context & Binary & RQ4 \\
    & ACL & Binary & RQ4 \\
    & Aggregation & Binary & RQ4 \\
    & Domain Events & Binary & RQ4 \\
    & Domain Services & Binary & RQ4 \\
    & Entities and Value Objects & Binary & RQ4 \\
  \hline
\end{tabular}
\caption{Data Extraction Form.}
\label{tab:data_extraction_form}
\end{table}

\subsubsection{Synthesis}
In this study, the data extraction process involved the collection of both qualitative and quantitative data from the final set of primary studies to derive our results. Specifically, for research questions (RQs) encompassing 1 to 4, we focused on extracting quantitative data. The frequencies and percentages of each identified category were then computed and reported to address these specific research questions. This approach allowed us to quantify and present the distribution and prevalence of different factors and concepts relevant to the research questions under consideration.

Conversely, for RQs pertaining to RQs 5 to 7, we concentrated on extracting qualitative data from the primary studies. To synthesize and interpret these qualitative findings, we applied a narrative synthesis method based on the guidance provided by Popay et al.~\cite{popay_guidance_2006}. This involved systematically extracting key information and important findings from each individual study. Subsequently, we conducted a comprehensive analysis and interpretation of the extracted qualitative data to draw overarching and meaningful conclusions in response to the respective research questions. The narrative synthesis approach allowed us to integrate diverse qualitative evidence, explore patterns and themes across the studies, and provide a cohesive and comprehensive understanding of the research questions in question.

\begin{sidewaystable}
\adjustbox{max height=\textheight, max width=\textwidth}{
\begin{tabular}{clcccc}
\hline
& Title & Author(s) & Year & Score \\
\hline
        \cite{singjai_patterns_2021} & Patterns on Deriving APIs and their Endpoints from Domain Models & Singjai et al. & 2021 & 4 \\
        \cite{maddodi_aggregate_2020} & Aggregate Architecture Simulation in Event-Sourcing Applications using Layered Queuing Networks & Maddodi et al. & 2020 & 7.5 \\
        \cite{peng_ianticorruption_2007} & IAnticorruption: a domain-driven design approach to more robust integration & Peng et al. & 2007 & 4.5 \\
        \cite{joselyne_partitioning_2018} & Partitioning microservices: a domain engineering approach & Josélyne et al. & 2018 & 5 \\
        \cite{perillo_daileon_2009} & Daileon: a tool for enabling domain annotations & Perillo et al. & 2009 & 4.5 \\
        \cite{le_jdomainapp_2019} & jDomainApp: A Module-Based Domain-Driven Software Framework & Le et al. & 2019 & 6.5 \\
        \cite{le_generating_2022} & Generating Multi-platform Single Page Applications: A Hierarchical Domain-Driven Design Approach & Le et al. & 2022 & 5 \\
        \cite{garbajosa_supporting_2018} & Supporting Large-Scale Agile Development with Domain-Driven Design & Uludağ et al. & 2018 & 7 \\
        \cite{wang_reference_2022} & A Reference Architecture for Blockchain-based Traceability Systems Using Domain-Driven Design and Microservices & Wang et al. & 2022 & 6.5 \\
        \cite{landre_architectural_2006} & Architectural improvement by use of strategic level domain-driven design & Landre et al. & 2006 & 6.5 \\
        \cite{kapferer_domain-driven_2020} & Domain-Driven Service Design: Context Modeling, Model Refactoring and Contract Generation & Kapferer et al. & 2020 & 5.5 \\
        \cite{camilli_actor-driven_2023} & Actor-driven Decomposition of Microservices through Multi-level Scalability Assessment & Camilli et al. & 2023 & 7 \\
        \cite{braun_tackling_2021} & Tackling Consistency-related Design Challenges of Distributed Data-Intensive Systems: An Action Research Study & Braun et al. & 2021 & 7 \\
        \cite{landre_agile_2007} & Agile enterprise software development using domain-driven design and test first & Landre et al. & 2007 & 6.5 \\
        \cite{braun_advanced_2021} & Advanced Domain-Driven Design for Consistency in Distributed Data-Intensive Systems & Braun et al. & 2021 & 6.5 \\
        \cite{zhao_applying_2021} & Applying Microservice Refactoring to Object-Oriented Legacy System & Zhao et al. & 2021 & 4 \\
        \cite{krause_microservice_2020} & Microservice Decomposition via Static and Dynamic Analysis of the Monolith & Krause et al. & 2020 & 5.5 \\
        \cite{snoeck_agile_2022} & Agile MERODE: a model-driven software engineering method for user-centric and value-based development & Snoeck et al. & 2022 & 6 \\
        \cite{skersys_domain_2012} & Domain Driven Development and Feature Driven Development for Development of Decision Support Systems & Danenas et al & 2012 & 4 \\
        \cite{hammoudi_domain-driven_2021} & Domain-Driven Architecture Modeling and Rapid Prototyping with Context Mapper & Kapferer et al. & 2021 & 6.5 \\
        \cite{hammarstrom_experience_2016} & Experience from integrating Domain Driven Software System Design into a Systems Engineering Organization & Hammarström et al. & 2016 & 5.5 \\
        \cite{da_silva_bpm2ddd_2022} & BPM2DDD: A Systematic Process for Identifying Domains from Business Processes Models & Da Silva et al. & 2022 & 6.5 \\
        \cite{hippchen_designing_2017} & Designing Microservice-Based Applications by Using a Domain-Driven Design Approach & Hippchen et al. & 2017 & 6 \\
        \cite{ding_enterprise_2020} & Enterprise service application architecture based on Domain Driven Model Design & Ding et al. & 2020 & 5 \\
        \cite{ozkan_refactoring_2023} & Refactoring with domain-driven design in an industrial context: An action research report & \"{O}zkan et al. & 2023 & 8 \\
        \cite{pereira_towards_2022} & Towards Transactional Causal Consistent Microservices Business Logic & Pereira et al. & 2022 & 6.5 \\
        \cite{joselyne_systematic_2021} & A Systematic Framework of Application Modernization to Microservice based Architecture & Joselyne et al. & 2021 & 4.5 \\
        \cite{wanderley_framework_2012} & A Framework to Diminish the Gap between the Business Specialist and the Software Designer & Wanderley et al. & 2012 & 5 \\
        \cite{koryl_active_2017} & Active resources concept of computation for enterprise software & Koryl et al. & 2017 & 4 \\
        \cite{mayer_approach_2018} & An Approach to Extract the Architecture of Microservice-Based Software Systems & Mayer et al. & 2018 & 5 \\
        \cite{vural_does_2021} & Does Domain-Driven Design Lead to Finding the Optimal Modularity of a Microservice? & Vural et al. & 2021 & 8 \\
        \cite{oukes_domain-driven_2021} & Domain-Driven Design applied to land administration system development: Lessons from the Netherlands & Oukes et al. & 2021 & 4.5 \\
        \cite{haser_is_2016} & Is business domain language support beneficial for creating test case specifications: A controlled experiment & Häser et al & 2016 & 6.5 \\
        \cite{le_domain_2018} & On domain driven design using annotation-based domain specific language & Le et al. & 2018 & 6 \\
        \cite{abeck_context_2019} & A Context Map as the Basis for a Microservice Architecture for the Connected Car Domain & Abeck et al. & 2019 & 4 \\
        \cite{bunder_model-driven_2019} & A model-driven approach for behavior-driven GUI testing & Bünder et al. & 2019 & 5.5 \\
\hline
\end{tabular}}
\caption{List of primary studies and their quality assessment scores.}
\label{tab:list_of_primary_studies}
\end{sidewaystable}

\section{Results} \label{sec:results}
This section presents a comprehensive analysis of the findings obtained from our SLR conducted on DDD and aims to address the research questions by providing a detailed exploration of the current state of research pertaining to DDD in the literature. Through a meticulous examination of various articles, studies, and sources, the section sheds light on the key insights and trends related to the implementation, effectiveness, and assessment of DDD in diverse software systems. By organizing the results around the research questions, this section offers valuable insights into the involvement of key stakeholders, the evaluation and measurement of DDD's effectiveness, and the overarching research landscape concerning this domain-centric architectural approach.

\subsection{(RQ1) What is the state of the art concerning DDD in the existing literature, in terms of study types, research venues, software systems, and the distribution between academic and industrial settings?} \label{subsec:rq1}
\begin{figure}[t]
  \centering
  \includegraphics[width=\linewidth]{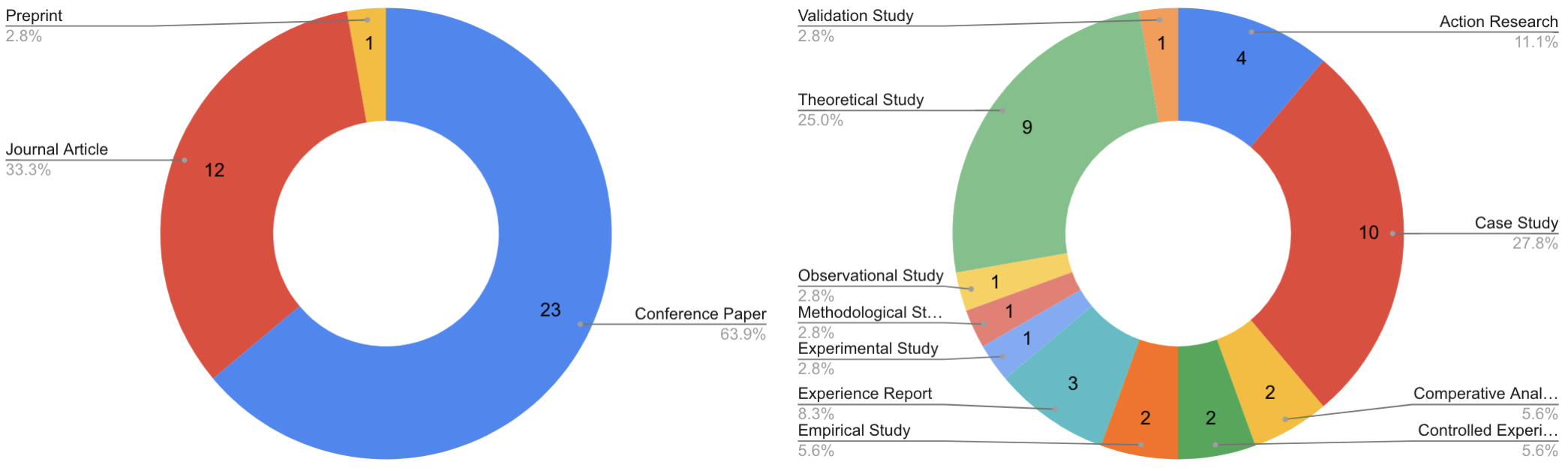}
  \caption{Breakdown of study types (left) and study designs (right).}
  \label{fig:breakdown_articletype_studytype}
\end{figure}

Based on our primary studies (36), the current state of research concerning DDD in the existing literature showcases studies conducted between 2006 and 2023. Among these studies, 23 (64\%) are \textit{Conference Papers}, 12 (33\%) are \textit{Journal Articles}, and 1 (3\%) study is from the \textit{Preprint} database, which also can be seen in Figure \ref{fig:breakdown_articletype_studytype}.

In terms of venues, a diverse range of conferences and journals have emerged as platforms for the dissemination of DDD research. These venues span the spectrum of software engineering and technology domains. Among them, the \textit{Conference on Object-Oriented Programming Systems, Languages, and Applications} takes a prominent role with 3 publications, showcasing its relevance in fostering discussions around object-oriented paradigms in software development. Additionally, the \textit{Symposium on Information and Communication Technology} have contributed 2 publications.

The most common study type is \textit{Case Study}, accounting for 10 (28\%) studies, followed by \textit{Theoretical Study} with 9 (25\%) studies. There are also 4 (11\%) \textit{Action Research} studies, 3 (8\%) \textit{Experience Reports}, 2 (6\%) \textit{Empirical Studies}, 2 (6\%) that are not explicitly mentioning a type but using empirical methodologies, \textit{Controlled Experiments}, 2 (6\%) \textit{Comparative Analyses}, 1 (3\%) \textit{Observational Study}, 1 (3\%) \textit{Methodological Study}, and 1 (3\%) \textit{Experimental Study}. The remaining studies were primarily conceptual, methodological, or design proposals with limited or no empirical validation.

In terms of measurement rigor, 17 studies did not report any concrete evaluation metrics. Those studies are theoretical or design-oriented. Many propose methods, tools, or architectural patterns without validating them through empirical means. While some are labeled as case studies or experience reports, their evidence is often anecdotal and lacks systematic measurement. This pattern suggests a gap in empirically grounded research in certain areas of DDD, particularly tool development and architectural modeling.

When cross-referencing these findings with the application domains, we observed that empirical evaluations were more likely to be found in industrial case studies involving microservices and distributed systems, while studies focused on tools or frameworks tended to lack rigorous empirical assessment. Furthermore, studies based on action research or industry collaborations generally provided more detailed insight into DDD's impact on software quality, maintainability, and organizational communication.

Regarding the software systems studied, \textit{Microservices} received the highest attention with 16 (44\%) studies, followed by \textit{Enterprise Software Systems} with 6 (17\%) studies, and DDD-related \textit{Tools} and \textit{System Agnostic} studies with 4 (11\%) studies each. Additionally, \textit{Standard Computer Applications} and \textit{Web Applications} were subjects of 2 (6\%) studies each, while \textit{Service-Oriented Applications} and \textit{Distributed Systems} each had 1 (3\%) study.

\begin{figure}[t]
  \centering
  \includegraphics[width=\linewidth]{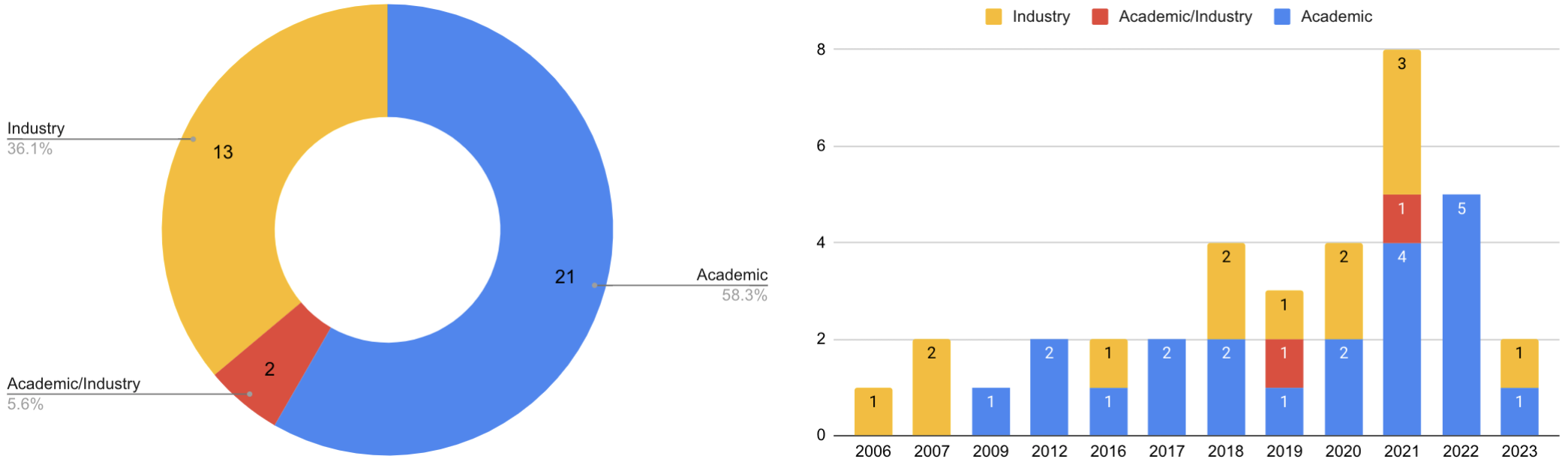}
  \caption{Breakdown of the primary studies based on their field of focus (left) and field of focus by publication year (right).}
  \label{fig:breakdown_field_years}
\end{figure}

Out of the 16 \textit{Microservices} studies, 10 (63\%) were conducted in an academic environment, while the remaining 6 (38\%) were performed in industrial settings. In contrast, \textit{Enterprise Software Systems }were predominantly studied in industry settings. Interestingly, DDD-related \textit{Tools} were exclusively implemented in academic settings.

The interest in DDD initially emerged in industry settings, as evidenced by the first 3 studies conducted between 2006 and 2007. Subsequently, the focus shifted to academic settings, with all 3 studies from 2009 to 2012 being conducted in academia. The trend then became more balanced, with an equal distribution of studies between academic and industrial settings in 2016.

\begin{figure}[t]
  \centering
  \includegraphics[width=\linewidth]{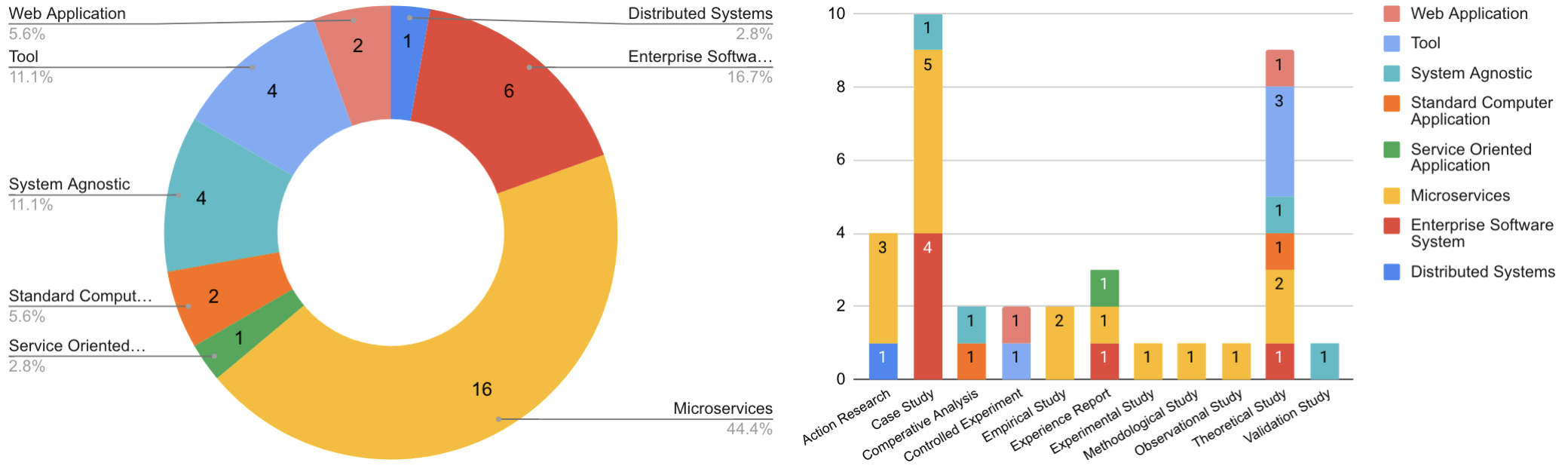}
  \caption{Breakdown of the primary studies based on applied software systems (left) and studies grouped by the study type and software systems (right).}
  \label{fig:breakdown_swsystem_studytypes}
\end{figure}

The distribution of studies across academic and industry settings reveals an interesting pattern. Out of the total 36 studies, 21 (58.3\%) were conducted in academia, indicating a dominant presence of academic research in this field. Conversely, 13 (36.1\%) studies were conducted in industry settings, and 2 (5.6\%) studies had a hybrid setting involving both academic and industry collaboration. This disparity in the academic-industry ratio suggests that DDD has garnered significant attention from the academic community, where researchers are actively exploring its concepts and applications. However, it also highlights the relevance and practical implications of DDD in industry settings, as evidenced by the notable number of industry-based studies. 

The existing literature demonstrates a diverse and evolving landscape of research on DDD, with a particular emphasis on Microservices and an increasing number of academic contributions. This suggests a growing recognition of the relevance and applicability of DDD in both academic and industrial contexts.

\subsection{(RQ2) What categories of software systems have demonstrated benefits from the implementation of DDD?} \label{subsec:rq2}
With this RQ, we present the categories of software systems that have demonstrated benefits from the implementation of DDD, based on the findings. The studies were analyzed, and their contributions to various software system categories were identified during the data extraction phase. The breakdown of the following data can also be visually seen in Figure \ref{fig:breakdown_swsystem_studytypes}.

\subsubsection{Microservices} \label{subsubsec:microservices}
Several studies have highlighted the advantages of applying DDD principles in the context of microservices architecture \cite{singjai_patterns_2021, joselyne_partitioning_2018, wang_reference_2022, camilli_actor-driven_2023, braun_advanced_2021, zhao_applying_2021, krause_microservice_2020, hippchen_designing_2017, ding_enterprise_2020, ozkan_refactoring_2023, pereira_towards_2022, joselyne_systematic_2021, mayer_approach_2018, vural_does_2021, oukes_domain-driven_2021, abeck_context_2019}. According to our learnings from those studies, by adopting DDD, organizations have been able to improve the modularity, maintainability, and scalability of microservices-based applications, enabling effective handling of complex and distributed systems. The use of DDD patterns and concepts in the design and development of microservices has been particularly beneficial in achieving clear boundaries between services, enabling better communication between development teams, and ensuring a strong alignment with business domains. For instance, Krause et al. \cite{krause_microservice_2020} applied DDD principles to modernize a real-world lottery system into a microservices-based architecture, demonstrating the approach's relevance in refactoring legacy systems in the public sector.

\subsubsection{Enterprise Software Systems}
Several studies have reported the benefits of applying DDD in the context of enterprise software systems \cite{maddodi_aggregate_2020, peng_ianticorruption_2007, garbajosa_supporting_2018, landre_agile_2007, skersys_domain_2012, koryl_active_2017}. The adoption of DDD principles has proven valuable in handling the increasing complexity of enterprise systems and addressing challenges related to legacy system modernization. By emphasizing bounded contexts and domain models, DDD enables more maintainable and flexible enterprise software architectures, allowing organizations to better adapt to changing business requirements.

\subsubsection{Web Applications}
Some studies have shown that DDD can be beneficial in the development of web applications \cite{le_generating_2022, bunder_model-driven_2019}. By applying DDD principles, web application developers have been able to improve the design and implementation of Single Page Applications (SPAs), achieving better cross-framework compatibility and enhancing productivity in SPA development.

\subsubsection{Tools and Standard Computer Applications}
Several studies have applied DDD in the context of tools \cite{perillo_daileon_2009, kapferer_domain-driven_2020, haser_is_2016} and standard computer applications \cite{le_jdomainapp_2019, hammarstrom_experience_2016}. DDD has been leveraged to improve code quality, enhance the software construction process, and provide more comprehensive solutions for domain modeling and software generation. In the case of tools, DDD has been employed to bridge the gap between business domain concerns and infrastructure, ensuring the maintainability and readability of code.

\subsubsection{Distributed Systems}
In this study, we classified systems based on the architectural types explicitly mentioned in the primary studies. However, some system types, such as Service-Oriented Architecture (SOA) (Section \ref{subsubsec:soa}) and Distributed Systems, may overlap in practice. For example, SOA is one possible way of realizing a distributed system.

In cases where the primary study did not specify the exact architectural type such as microservices or SOA, but specifies it is a distributed system, we used the more generic term Distributed Systems to accurately reflect the nature of the system without over-specification.

This classification ensures that we do not make assumptions about the specific system architecture when such information is not available, and it provides a comprehensive view of how DDD is applied in various types of distributed systems. We acknowledge that system types like SOA and Distributed Systems are not always mutually exclusive, and we carefully categorized the systems according to the information available in the primary studies.

This approach was applied to Braun et al.'s study \cite{braun_tackling_2021}, where it reported that DDD has demonstrated benefits in the development of distributed systems. By controlling concurrent data access and achieving semantic compatibility in domain models, DDD has been instrumental in tackling consistency-related design challenges in distributed data-intensive systems. Similarly, Wang et al. \cite{wang_reference_2022} utilized DDD to architect a blockchain-based traceability system in the global supply chain domain, where managing bounded contexts across distributed actors was critical.

\subsubsection{Service-Oriented Applications (SOA)} \label{subsubsec:soa}
Landre et al. \cite{landre_architectural_2006} use DDD patterns to improve the software architecture of a large enterprise system, which is designed as a SOA. The authors mention that establishing a DDD context map helped in scoping the project and identifying architectural problems such as complex dependencies and gateway roles.

In this study, we distinguish between SOA and Microservices as separate architectural classifications, not to make a strict ontological separation, but to reflect the terminology used in the primary studies themselves. While both paradigms promote modular and service-based decomposition, they also differ in typical implementations: microservices emphasize lightweight, independently deployable components with decentralized governance, whereas SOA is often realized with centralized orchestration and heavier protocols (e.g., ESB and SOAP) \cite{dragoni_microservices_2017, fritzsch_microservices_2019}.

We acknowledge that this boundary is not universally agreed upon, and many researchers recognize significant overlap between SOA and microservices. As noted by Lewis and Fowler \cite{lewis2014microservices}, the distinction lies more in implementation philosophy than fundamental principles. Our categorization aims to reflect how the systems were described and studied in the reviewed literature, without implying that SOA and microservices are categorically distinct. Future work could explore how these architectural paradigms co-evolve in practice.

\subsubsection{System Agnostic}
Certain studies have shown that DDD can provide valuable contributions for the implementation of software which is suitable across various system categories \cite{snoeck_agile_2022, hammoudi_domain-driven_2021, da_silva_bpm2ddd_2022, wanderley_framework_2012}. In these studies, DDD principles and patterns have been applied to implement practitioner frameworks and methods to address communication gaps between business specialists and software designers, improve software system modularity, and support the creation of context maps. 

\subsubsection{DDD Tools}
DDD has been applied to improve code quality, readability, and maintainability through the development of supportive tooling. These tools fall into two main categories. Annotation-based DDD frameworks, such as Daileon by Perillo et al. \cite{perillo_daileon_2009}, aim to separate business domain concerns from infrastructure code. In contrast, DDD modeling tools, such as Context Mapper by Kapferer et al. \cite{kapferer_domain-driven_2020}, provide higher-level support for visualizing and evolving Context Maps and other domain artifacts.

\subsection{(RQ3) Which specific software development concerns have been addressed through the application of DDD in previous studies?} \label{subsec:rq3}
The SLR conducted in this study sought to identify the specific software development concerns that have been addressed through the application of DDD in previous research. The results revealed several challenges and motivations encountered in various software development projects, encompassing diverse domains such as microservices architecture, legacy system modernization, distributed systems, and complex decision support systems. The subsequent information present a comprehensive overview of the identified challenges and motivations for adopting DDD.

\subsubsection{Addressing API Design Problems in Distributed Systems}
One of the prominent challenges addressed by DDD in software development projects is related to API design in distributed systems, particularly in microservice-based systems. This challenge revolves around missing guidance on how to derive APIs and API endpoints from domain model elements. The primary concern is that coupling smells lead to poor API design in distributed systems. These issues result in shallow APIs that are difficult to understand, maintain, and evolve. Singjai et al. \cite{singjai_patterns_2021} propose clear guidance on how to derive APIs
and API endpoints from domain model elements using DDD patterns.

\subsubsection{Investigating Performance in Component-Based Software}
Command Query Responsibility Segregation (CQRS) and Event Sourcing (ES) are two architectural patterns often used in conjunction with DDD to improve scalability and traceability. CQRS separates write operations (commands) from read operations (queries), enabling systems to optimize each independently. Event Sourcing, on the other hand, persists all changes to the application state as a sequence of immutable events rather than storing the current state directly. In this approach, the current state is reconstructed by replaying these events. These patterns align well with DDD’s tactical building blocks—such as aggregates and domain events—and have been leveraged in studies to evaluate performance characteristics under different workload patterns \cite{maddodi_aggregate_2020}.

In pursuit of achieving optimized performance and resource metrics in component-based software, DDD has been used in research to investigate the effect of workload patterns. Specifically, the research aimed to gain insight into how patterns like CQRS and ES are created, thus providing valuable guidelines for architects to select appropriate architectures based on workload characteristics \cite{peng_ianticorruption_2007}.

\subsubsection{Reducing Complexity in Currency Exchange Systems}
A prevalent software development problem addressed by DDD pertains to reducing complexity and high coupling in currency exchange systems. The case in question involves the integration layer between a new currency exchange system (SPOT) and a legacy back office system (TBS), wherein the integration layer (TBSExport) had become an unwieldy burden with unpredictable side effects following minor changes. To tackle this complexity and streamline the integration process, researchers sought to devise effective solutions using DDD principles \cite{peng_ianticorruption_2007}.

\subsubsection{Microservice Size, Partitioning and Defining API Endpoints} 
Microservice architecture introduces the challenge of determining the appropriate size of individual microservices, thereby necessitating a conceptual methodology for partitioning microservices based on domain engineering techniques. Improperly sized microservices can lead to complexity management issues, where too large services become monolithic and too small services create an overly complex system. This can also result in performance bottlenecks and scalability challenges, as well as data consistency problems due to poorly coordinated related data spread across services. Additionally, undefined or poorly defined boundaries complicate maintenance and evolution, and misaligned microservice boundaries with team boundaries hinder team autonomy and require excessive coordination. DDD has been instrumental in addressing this challenge, aiming to achieve optimal microservice design and composition within the context of a microservices architecture \cite{singjai_patterns_2021, joselyne_partitioning_2018}.

\subsubsection{Improving Code Quality with Domain Annotations}
In the pursuit of enhancing code quality, readability, and maintainability while ensuring a clear separation between the business domain and infrastructure concerns, DDD has been leveraged to address challenges associated with using annotations in DDD. The proposed concept of "domain annotations" by Perillo et al. \cite{perillo_daileon_2009} serves to mitigate these challenges and fosters a robust software design approach.

\subsubsection{Enhancing Software Construction and Development Environment}
The lack of generative, module-based software construction and seamless development environment integration poses significant challenges in software development. Without these capabilities, developers face difficulties in automating the generation of software from domain models, leading to increased manual effort and potential for errors \cite{le_jdomainapp_2019}. This can result in inconsistencies between the domain model and the implemented software, hindering maintainability and scalability. Additionally, the absence of integrated development tools means that developers must switch between different environments and tools, disrupting their workflow and reducing productivity. By addressing these issues, in the work of Le et al. \cite{le_jdomainapp_2019} the adoption of the jDomainApp framework and an Eclipse IDE plugin has been proposed as a solution to these challenges, thereby optimizing software construction and development processes.

\subsubsection{Multi-Platform SPA Development}
The challenges faced by Single Page Application (SPA) developers encompass two crucial aspects: designing SPAs that can effectively work across different SPA frameworks (e.g., Angular, React, React Native, and Vue.js) and translating this design into an intermediate high-level language that can seamlessly transform into a target framework of choice. DDD offers a multi-platform, hierarchical approach to tackle these challenges, significantly enhancing SPA development productivity \cite{le_generating_2022}.

\subsubsection{Scaling Agility in Large Organizations}
Large organizations encounter a myriad of challenges when striving to scale agility, including effective coordination and communication between agile teams, inter-team dependencies, and the need for clearly defined requirements. By adopting DDD, enterprises have sought to address these challenges and provide architectural guidance to support large-scale agile development. However, it is important to acknowledge that successful DDD implementation in large organizations necessitates experienced enterprise architects to provide effective guidance \cite{garbajosa_supporting_2018}.

\subsubsection{Complexities in Blockchain-Based Traceability Systems (BTS)}
BTSs are software systems designed to ensure transparent, tamper-proof tracking of products or data throughout a supply chain or process. These systems leverage blockchain technology to record transactions and state changes in an immutable ledger, enhancing trust, provenance, and accountability.

In the context of DDD, BTSs benefit from patterns such as Bounded Contexts and Aggregates to model complex interactions and responsibilities across different stakeholders and subsystems. As reported in \cite{wang_reference_2022}, the application of DDD in BTS improves the maintainability and extensibility of these systems by clearly delineating domain boundaries and aligning technical implementations with business processes.

The design and development of BTSs pose unique challenges and complexities, including domain model intricacies, modeling boundaries, and managing the relationship between different bounded contexts. Researchers have employed DDD and microservice architecture to address these challenges, thereby improving the cohesiveness, maintainability, and extensibility of BTSs in handling complex data in the global supply chain in \cite{wang_reference_2022}.

\subsubsection{Challenges in Model Decomposition}
Decomposing software systems into cohesive and loosely coupled modules poses significant challenges for software engineers and service designers. DDD patterns have been instrumental in addressing this challenge, offering a clear and concise interpretation of these patterns, despite debates and differing interpretations among practitioners \cite{hammoudi_domain-driven_2021}.

\subsubsection{Legacy System Modernization and Refactoring}
Legacy systems often suffer from outdated technologies, poor documentation, and a lack of alignment between the system's current state and the evolving business requirements. This misalignment creates a communication gap between business specialists, who understand the domain requirements, and software designers, who are tasked with implementing these requirements \cite{ozkan_refactoring_2023}. As a result, changes and enhancements become risky, time-consuming, and error-prone. DDD addresses these issues by providing a structured approach to align the domain model with business needs, facilitating better communication and understanding. This alignment enables more effective refactoring and modernization efforts, ensuring that the updated system meets current business requirements while maintaining or improving system integrity and performance \cite{ozkan_refactoring_2023, wanderley_framework_2012}.

\begin{table}[ht]
\centering
\adjustbox{max width=\textwidth}{
\begin{tabular}{p{3.5cm} | p{10.5cm}}
\toprule
\textbf{Software System Category} & \textbf{Reported Benefits of DDD} \\
\midrule
\textbf{Microservices} & 
Improved modularity, maintainability, and scalability; better service boundary definition; enhanced team autonomy; reduced coupling through bounded contexts and context mapping \cite{singjai_patterns_2021, joselyne_partitioning_2018, wang_reference_2022, camilli_actor-driven_2023}. \\
\hline
\textbf{Enterprise Software Systems} & 
Support for legacy system modernization; clearer architecture via bounded contexts; improved adaptability to evolving business requirements \cite{maddodi_aggregate_2020, garbajosa_supporting_2018, landre_agile_2007}. \\
\hline
\textbf{Web Applications} & 
Better SPA design across frameworks; productivity boost through DDD-driven component generation \cite{le_generating_2022, bunder_model-driven_2019}. \\
\hline
\textbf{Tools} & 
Improved code quality and modularity through domain annotations; enhanced maintainability by separating domain logic from infrastructure concerns \cite{perillo_daileon_2009, kapferer_domain-driven_2020}. \\
\hline
\textbf{Distributed Systems} & 
Addressed consistency challenges; improved semantic compatibility and isolation between modules \cite{braun_tackling_2021}. \\
\hline
\textbf{SOA / Service-Oriented Applications} & 
Clear project scoping; identification of architectural problems (e.g., dependency complexity) using context maps \cite{landre_architectural_2006}. \\
\hline
\textbf{System-Agnostic Approaches} & 
Support for agile methods; communication alignment between business and technical stakeholders; improved modeling modularity \cite{snoeck_agile_2022, da_silva_bpm2ddd_2022}. \\
\bottomrule
\end{tabular}
}
\caption{Reported benefits of DDD across different software system categories, as identified in the primary studies.}
\label{tab:ddd_benefits_by_system}
\end{table}

\subsection{(RQ4) What are the most common DDD patterns utilized in previous research, such as ACLs, Bounded Contexts, and other related approaches?} \label{subsec:rq4}
\begin{figure}[t]
  \centering
  \includegraphics[width=\linewidth]{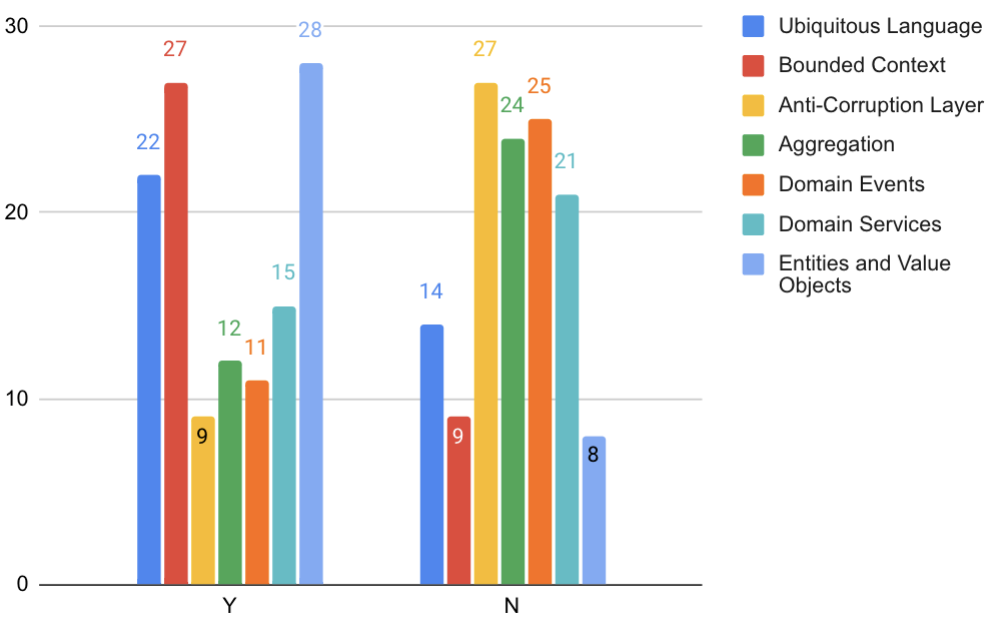}
  \caption{Breakdown of the applied DDD patterns in the primary studies. Applied DDD patterns marked with "Y" and not applied pattern marked with "N".}
  \label{fig:ddd_patterns_breakdown_chart}
\end{figure}

In this SLR on DDD, we investigated the most commonly utilized DDD patterns across various software systems. The analysis was based on a dataset comprising studies identified through our study selection process explained in Section \ref{subsec:primary_study_identification_and_selection}. Each study was examined for the presence or absence of specific DDD patterns, marked by "Y" for used and "N" for not used patterns, during our data extraction process. Our results are also visually shown in Figure \ref{fig:ddd_patterns_breakdown_chart}.

The results reveal that certain DDD patterns are prevalent in the majority of the analyzed studies, indicating their significance and widespread adoption in the DDD context. The following DDD patterns were found to be most commonly utilized:

\begin{itemize}
    \item[1.] \textit{Ubiquitous Language.} Ubiquitous language is a fundamental DDD pattern, and our findings indicate that it is frequently employed in various software systems, including Microservices, Enterprise Software Systems and Web Applications \cite{maddodi_aggregate_2020, peng_ianticorruption_2007, joselyne_partitioning_2018, le_generating_2022, garbajosa_supporting_2018, krause_microservice_2020, skersys_domain_2012, hammoudi_domain-driven_2021, ozkan_refactoring_2023, pereira_towards_2022, joselyne_systematic_2021, oukes_domain-driven_2021}.
    \item[2.] \textit{Bounded Context.} Bounded contexts were found to be widely utilized in Microservices, Enterprise Software Systems and Web Applications \cite{maddodi_aggregate_2020, peng_ianticorruption_2007, joselyne_partitioning_2018, le_generating_2022, garbajosa_supporting_2018, camilli_actor-driven_2023, braun_tackling_2021, braun_advanced_2021, zhao_applying_2021, snoeck_agile_2022, hammoudi_domain-driven_2021, ozkan_refactoring_2023, pereira_towards_2022}.
    \item[3.] \textit{Entities and Value Objects.} Entities and value objects, representing domain concepts, were extensively adopted in Microservices, Enterprise Software Systems and Web Applications \cite{maddodi_aggregate_2020, peng_ianticorruption_2007, joselyne_partitioning_2018, le_generating_2022, garbajosa_supporting_2018, wang_reference_2022, braun_tackling_2021, snoeck_agile_2022, skersys_domain_2012, hammoudi_domain-driven_2021, hammarstrom_experience_2016, da_silva_bpm2ddd_2022, hippchen_designing_2017, ozkan_refactoring_2023, pereira_towards_2022, joselyne_systematic_2021, wanderley_framework_2012}.
\end{itemize}

In contrast, the \textit{ACL} was identified as the least used DDD pattern across different software systems. Several studies did not utilize the ACL in their respective projects \cite{peng_ianticorruption_2007, le_jdomainapp_2019, landre_architectural_2006, kapferer_domain-driven_2020, koryl_active_2017, haser_is_2016, bunder_model-driven_2019}.

The \textit{ACL} serves as a means to protect the integrity of the main domain from external and legacy systems, and its limited adoption suggests that certain software projects may have addressed integration challenges through alternative approaches or may not have encountered a specific need for this pattern.

While Ubiquitous Language, Bounded Context, and Entities/Value Objects were widely observed across studies, patterns such as Core Domain and Shared Kernel were rarely reported. Only a limited number of studies (e.g., \cite{landre_architectural_2006, ozkan_refactoring_2023}) referred to the Core Domain explicitly, often in the context of prioritizing critical business logic. Shared Kernel was not discussed as a standalone concept in any study, although collaboration between teams was sometimes implied in studies using Bounded Contexts. Similarly, upstream–downstream relationships—central to context mapping in Strategic Design—were briefly mentioned in a few cases but not analyzed in depth.

\subsection{(RQ5) What are the challenges encountered during the implementation of DDD in software development projects?} \label{subsec:rq5}
The challenges encountered during the implementation of DDD in software development projects are diverse and demand careful consideration to ensure successful adoption. In this RQ, we present the key challenges identified through primary studies. To support the narrative discussion, we also summarize the main DDD implementation challenges and mitigation strategies identified across the primary studies. Table \ref{tab:ddd_challenges_mitigations} presents a concise mapping that highlights both the barriers encountered and the approaches taken to address them in practice.

\subsubsection{Addressing Design Complexities}
Several studies highlight the complexity associated with implementing DDD patterns, particularly in the context of microservices and distributed systems. The integration layer between different Bounded Contexts can become burdensome, leading to high coupling and unpredictability in the system. This challenge, for instance, is exemplified in the case of the currency exchange system (SPOT) and the legacy back-office system (TBS) in Peng et al. \cite{peng_ianticorruption_2007}. Finding the right balance between high cohesion within services and loose coupling between them is crucial for scalability and maintainability \cite{peng_ianticorruption_2007, joselyne_partitioning_2018}.

\begin{table}[ht]
\centering
\adjustbox{max width=\textwidth}{
\begin{tabular}{p{5.2cm} | p{10.5cm}}
\toprule
\textbf{Challenge} & \textbf{Mitigation Strategy and Concrete Techniques (from Literature)} \\
\midrule
Design complexity in distributed systems & Use of context mapping and bounded contexts to reduce coupling; ACLs to encapsulate legacy interactions \cite{peng_ianticorruption_2007, joselyne_partitioning_2018} \\
\hline
Defining microservices boundaries & Domain decomposition using business-aligned subdomains and context maps; Use of design workshops to iteratively refine service granularity \cite{joselyne_partitioning_2018, hippchen_designing_2017, vural_does_2021} \\
\hline
Managing complex domain models & Collaborative modeling with domain experts; event storming sessions; maintaining cohesion through tactical DDD patterns \cite{da_silva_bpm2ddd_2022, kapferer_domain-driven_2020, wang_reference_2022} \\
\hline
Integration with legacy systems & Use of ACLs to decouple legacy dependencies and allow incremental migration \cite{peng_ianticorruption_2007} \\
\hline
Communication gaps between domain experts and developers & Establishing Ubiquitous Language and domain co-design practices; continuous involvement of stakeholders \cite{wanderley_framework_2012, kapferer_domain-driven_2020} \\
\hline
Model-code misalignment in DSL-based tooling & Use of tools like Context Mapper for synchronized modeling and code generation; contract-based validation \cite{kapferer_domain-driven_2020} \\
\hline
Onboarding and lack of DDD expertise & Internal DDD training sessions; mentoring by experienced architects; gradual learning curve through pilot projects \cite{ozkan_refactoring_2023, hippchen_designing_2017} \\
\hline
Pattern interpretation disagreements & Team workshops to align on DDD pattern usage; creation of design documentation to guide consistent practices \cite{hammoudi_domain-driven_2021, ozkan_refactoring_2023} \\
\hline
Limited availability of domain experts & Early involvement of domain specialists; use of domain walkthroughs and business process models to extract knowledge \cite{vural_does_2021, da_silva_bpm2ddd_2022} \\
\bottomrule
\end{tabular}}
\caption{Challenges in implementing DDD and corresponding mitigation strategies and techniques reported in the literature.}
\label{tab:ddd_challenges_mitigations}
\end{table}

\subsubsection{Identifying Optimal Microservices Boundaries}
Decomposing monolithic systems into microservices presents its own set of challenges. Determining the appropriate size and boundaries of microservices is not straightforward and requires careful consideration of the domain and workload characteristics \cite{joselyne_partitioning_2018}. The granularity of microservices can be subjective, leading to potential disagreements among development teams \cite{hammoudi_domain-driven_2021}. Ensuring that microservices align with business needs and subdomains can be challenging \cite{hippchen_designing_2017}. Additionally, finding suitable examples that demonstrate DDD-driven microservices architecture is often noted as a limitation \cite{vural_does_2021}.

\subsubsection{Managing Domain Model Complexity}
When adopting DDD, managing the complexity of the domain model becomes a crucial challenge. Modeling boundaries between different Bounded Contexts and ensuring the relationships between them are properly managed require deep knowledge about the organization's domain \cite{da_silva_bpm2ddd_2022}. The complexity of the domain model and the interactions between bounded contexts can hinder the process of abstraction and generalization of elements \cite{da_silva_bpm2ddd_2022}. This challenge is especially pronounced when dealing with large-scale and complex data-intensive systems \cite{wang_reference_2022}.

\subsubsection{Integrating DDD with Existing Frameworks}
Integrating DDD into existing software frameworks can present compatibility issues, as DDD introduces encapsulation and abstraction levels that may not align seamlessly with existing structures \cite{ozkan_refactoring_2023}. Refactoring in an actively developed codebase can also prove challenging, especially when engineers lack prior experience with DDD \cite{ozkan_refactoring_2023}.

\subsubsection{Communication Gap between Business Specialists and Software Designers}
Incorporating DDD requires effective communication between business specialists and software designers to ensure consistency in requirements and conceptual models \cite{wanderley_framework_2012}. Misalignment in communication can lead to discrepancies in the final design and implementation.

\subsubsection{Model-Code Gap in Domain-Specific Language (DSL)}
A domain-specific language (DSL) is a computer language specialized to a particular application domain \cite{evans_domain-driven_2004}. Studies that focus on DSL-based DDD implementations point out potential weaknesses, including the existence of a "model-code" gap \cite{kapferer_domain-driven_2020}. The lack of a deeper knowledge about the domain may hinder the process of abstraction and generalization, leading to artifacts that may not align with the organization's domain \cite{da_silva_bpm2ddd_2022}.

\subsubsection{Impact of Developer's Experience and Challenges in Onboarding}
The level of experience and familiarity with DDD concepts among the development team can influence the effectiveness of DDD implementation \cite{hippchen_designing_2017}. Developers with prior experience in DDD are more likely to understand and apply DDD patterns effectively, resulting in better-designed and more maintainable software systems \cite{hippchen_designing_2017}. On the other hand, inexperienced developers may struggle with grasping the complexities of DDD and may not fully utilize the benefits of the approach \cite{hippchen_designing_2017}. Also, introducing DDD to a development team that has little or no prior exposure to DDD can be a challenge \cite{ozkan_refactoring_2023}. Onboarding developers and providing the necessary training and resources to understand DDD concepts and principles can require additional time and effort \cite{ozkan_refactoring_2023}. This challenge can be particularly pronounced in organizations that are transitioning from traditional software development approaches to DDD.

\subsubsection{Navigating Controversial Debates}
Controversial debates and differing interpretations exist among practitioners regarding the application and combination of DDD patterns \cite{hammoudi_domain-driven_2021, ozkan_refactoring_2023}. Developers may face challenges in deciding which patterns to adopt and how to effectively integrate them into the development process. This can result in disagreements within the team and may require additional effort to reach a consensus.

\subsubsection{Need for Experienced Domain Experts}
Incorporating DDD requires a deep understanding of the domain and business requirements. Experienced domain experts are valuable in guiding the development team to create accurate and effective domain models \cite{vural_does_2021}. However, the availability of such domain experts may be limited, especially in specialized domains, and this can hinder the DDD adoption process.

\subsection{(RQ6) How has the effectiveness of DDD been assessed and measured in prior studies?} \label{subsec:how_effectiveness_measured}
The effectiveness of DDD has been assessed and measured in prior studies through various evaluation methods and metrics. It is evident from the data that prior studies have employed a combination of quantitative and qualitative evaluation methods to assess the effectiveness of DDD. These methods are described in more detail in the following subsections.

\subsubsection{Performance and Resource Metrics} 
In the case of Maddodi et al. \cite{maddodi_aggregate_2020} the effectiveness of DDD was measured by evaluating the performance and resource metrics of the software system using Command Query Responsibility Segregation (CQRS) and Event Sourcing (ES) frameworks. The study reported relative error values for throughput and response time for different scenarios.

\subsubsection{Feedback from Stakeholders} In \cite{garbajosa_supporting_2018} the effectiveness of DDD practices, such as event storming workshops and domain modeling, was measured through the adoption and use of these practices by agile teams. The study also considered feedback from stakeholders, including Program Managers, Product Owners, and Domain Architects, on the benefits and value of DDD in supporting large-scale agile software development.

\subsubsection{Context Mapping and Complexity Reduction} Landre et al. \cite{landre_architectural_2006} evaluated the effectiveness of DDD, particularly the strategic part of it, by using context mapping and context relationships to analyze the software architecture and identify complexities. The study demonstrated how DDD provides mechanisms to address the identified problems and improve the quality of the Enterprise Architecture and derived software architectures by applying responsibility layers and improving scoping of projects.

\subsubsection{Effectiveness of Design Guidelines} Braun et al. \cite{braun_tackling_2021} conducted an assessment to determine the efficacy of design guidelines pertaining to distributed data-intensive systems. The evaluation encompassed various metrics, including the proportion of compatible domain operations and the extent of trivial aggregates within the domain model. 

\subsubsection{Questionnaires and Developer Perception} The study by Ozkan et al. \cite{ozkan_refactoring_2023}, the efficacy of DDD was evaluated using a combination of qualitative methods, including focus group sessions, semi-structured interviews, and think-aloud protocols. Additionally, a quantitative analysis was conducted using the Technology Acceptance Model (mTAM) questionnaire. The findings revealed that DDD positively impacted software maintainability, as developers perceived it. Similarly, Braun et al. \cite{braun_tackling_2021} undertook an assessment of design guidelines effectiveness by gauging developers' perceptions in their study. 

\subsubsection{Measuring Effectiveness of ECD3 Guidelines} In Braun et al. \cite{braun_advanced_2021} the effectiveness of the DDD measured based on Eventually Consistent Domain-Driven Design Guidelines (ECD3),  assessed by conducting a comparative analysis between compatible and incompatible domain operations in the redesigned domain models. The implementation of the recommended ECD3 guidelines exhibited a notable enhancement in the proportion of compatible operations within the domain models.

\subsubsection{Software Generation and Complexity Analysis} The effectiveness of the DDD-based software generation method proposed in \cite{le_jdomainapp_2019} was evaluated by examining its linear time complexity and scalability on real-world software of considerable magnitude, through the utilization of the jDomainApp framework.

\subsubsection{Time and Resource Consumption} In \cite{haser_is_2016} the effectiveness of including business domain concepts in the DSL was measured by evaluating the perceived quality, creation time, and length of test case specifications.

\subsubsection{Granularity and Coupling Analysis} The study \cite{vural_does_2021} assessed the effectiveness of DDD in microservices by comparing the coupling and cohesion values of different examples. The study concluded that original examples achieved a good balance between low coupling and acceptable cohesion.

\subsubsection{Prototype Application} In \cite{krause_microservice_2020} the effectiveness of the approach for modernizing legacy lottery applications into microservices was evaluated through its successful application to a real-world lottery application.

\subsubsection{Metamodel and Design Models} Le et al. \cite{le_generating_2022} aimed to assess the efficacy of DDD in constructing Single Page Applications (SPAs) compatible with multiple platforms. For this purpose, they developed a novel metamodel and implemented a hierarchical approach within the DDD framework. The outcomes of the study revealed that employing the domain model as a central component and adhering to the hierarchical DDD approach yielded favorable design models alongside prosperous generation of SPAs tailored for various platforms.

\subsection{(RQ7) Who are the key stakeholders involved and benefited in the implementation of DDD?} \label{subsec:rq7}

In the implementation of DDD, several key stakeholders are involved to ensure its successful application. These stakeholders encompass a diverse set of roles, including software engineers, architects, project managers, and domain experts. Through our SLR, we have identified the following key stakeholders.

\subsubsection{Software Engineers} Software Engineers play a central role in implementing DDD principles and practices in various software systems. They are responsible for translating the domain model into code, implementing domain-specific logic, and ensuring the alignment of the software architecture with the domain requirements \cite{singjai_patterns_2021, maddodi_aggregate_2020, peng_ianticorruption_2007, snoeck_agile_2022, ozkan_refactoring_2023}.

\subsubsection{Architects} Architects are crucial in the implementation of DDD as they provide strategic guidance and oversight on the overall software design. They collaborate with Software Engineers and other stakeholders to define the high-level architecture, bounded contexts, and domain models that capture the domain knowledge effectively \cite{maddodi_aggregate_2020, hammoudi_domain-driven_2021, da_silva_bpm2ddd_2022}.

\subsubsection{Project Managers} Project managers play a vital role in coordinating DDD implementation efforts. They ensure that DDD practices align with project goals, allocate resources appropriately, and facilitate effective communication between different teams and stakeholders \cite{peng_ianticorruption_2007, garbajosa_supporting_2018}.

\subsubsection{Domain Experts} Domain experts are individuals who possess in-depth knowledge of the specific business domain for which the software system is being developed. Their active involvement is critical for refining the domain model, identifying key domain concepts, and validating the software's alignment with real-world domain requirements \cite{singjai_patterns_2021, garbajosa_supporting_2018, da_silva_bpm2ddd_2022}.

In addition to the primary stakeholders mentioned above, the literature also identifies other relevant participants involved in the implementation of DDD. These participants include Quality Assurance (QA) Engineers, DevOps Engineers, UX Designers, Business Analysts, Systems Analysts, Data Modeling Specialists, and end-users \cite{peng_ianticorruption_2007, braun_advanced_2021, ozkan_refactoring_2023, oukes_domain-driven_2021, haser_is_2016, bunder_model-driven_2019}.

Taking our findings in consideration, we identified that the successful implementation of DDD requires close collaboration and communication among these diverse stakeholders to ensure that the developed software system effectively represents the domain's complexity and meets the business needs. The active engagement of software engineers, architects, project managers, and domain experts, along with the contributions from other relevant participants, enables the application of DDD principles to create robust and domain-centric software solutions.

\begin{figure}[t]
  \centering
  \includegraphics[width=\linewidth]{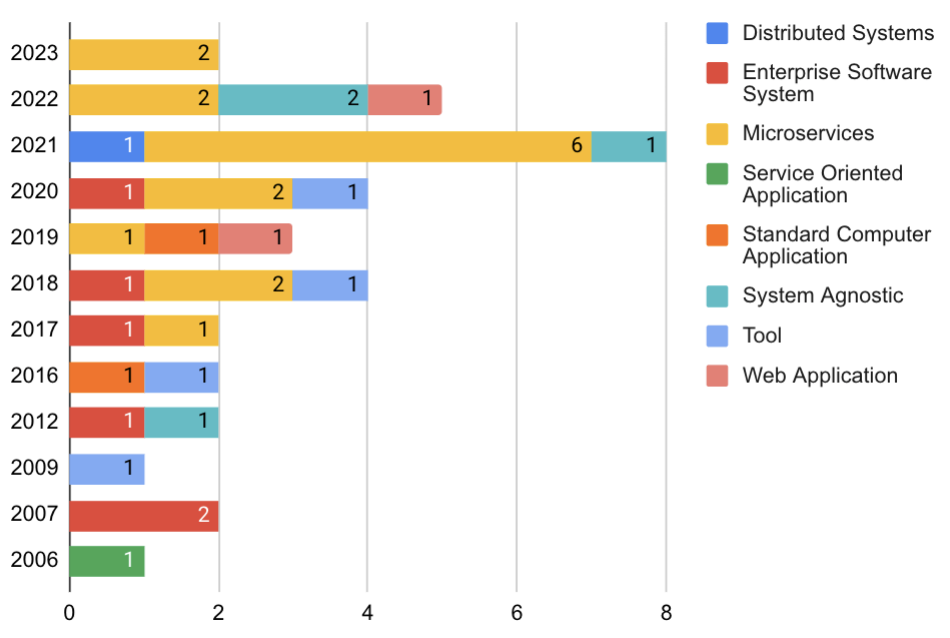}
  \caption{DDD studies grouped by their software systems and publication years.}
  \label{fig:publication_year_vs_software_systems}
\end{figure}

\section{Discussions}
In this section, we aim to consolidate and evaluate the extensive range of information obtained through our RQs in order to present a comprehensive overview of the effectiveness, challenges, and potential consequences associated with the adoption of DDD in software development. Throughout this section, we will do an examination of the findings by establishing connections across various studies and identifying recurring themes and patterns. It is important to acknowledge that certain aspects of the discussion have already been addressed in Section \ref{sec:results} when answering to the RQs. Nevertheless, within this dedicated section, we will explore further nuances, ramifications, and potential applications attributable to DDD based on the available evidence.

\subsection{Adoption and Implementation of DDD}
Several studies report successful implementations of DDD, showcasing its ability to improve software design, maintainability, scalability, and agility. DDD concepts, such as Ubiquitous Language, Bounded Contexts, Aggregates, Entities, Value Objects, and Domain Events, are commonly utilized to model complex domains effectively. Many researchers emphasize the significance of defining clear domain boundaries and using a common language between domain experts and software designers to ensure effective communication and understanding. Additionally, DDD is frequently integrated with other software development paradigms, such as microservices and event-driven architecture, to achieve better modularity and maintainable systems.

\subsection{Ubiquitous Language as a Core Principle of DDD}
The data analyzed points out that the Ubiquitous Language stands as a foundational principle of DDD. It ensures a shared and consistent understanding of the domain between stakeholders, including business specialists, software designers, and developers \cite{evans_domain-driven_2004}. The studies consistently emphasize the importance of Ubiquitous Language in facilitating communication, clarifying domain concepts, and improving collaboration among team members \cite{singjai_patterns_2021, maddodi_aggregate_2020, joselyne_partitioning_2018, perillo_daileon_2009, le_jdomainapp_2019, le_generating_2022, garbajosa_supporting_2018, skersys_domain_2012, hippchen_designing_2017, ding_enterprise_2020, ozkan_refactoring_2023, pereira_towards_2022, wanderley_framework_2012, le_domain_2018}.

\begin{table}[t]
  \centering
  \begin{tabular}{cl}
    \hline
    \ & Strengths \\
    \hline
    S1 & Aligns software systems with business domains conceptually. \\
    S2 & Widespread use of tactical/strategic patterns (e.g., Ubiquitous \\
       & Language, Bounded Context, Entities, Aggregates). \\
    S3 & Demonstrated benefits in modularity and maintainability, \\
       & especially in microservices. \\
    \hline
    \ & Weaknesses \\
    \hline
    W1 & Limited empirical evaluation in several studies. \\
    W2 & Underreporting of strategic patterns (e.g., Shared Kernel, \\
       & Core Domain, upstream/downstream relations). \\
    W3 & High learning curve for newcomers to DDD. \\
    \hline
    \ & Opportunities \\
    \hline
    O1 & Growing interest in academia and industry since 2017. \\
    O2 & Potential for stronger empirical and comparative research. \\
    O3 & Increasing tool support (e.g., DSLs, context-mapping tools). \\
    \hline
    \ & Threats \\
    \hline
    T1 & Risk of superficial or inconsistent DDD adoption. \\
    T2 & Conceptual disagreements and ambiguity in implementation. \\
    T3 & Dependence on domain experts may limit broader adoption. \\
    \hline
  \end{tabular}
  \caption{SWOT analysis of DDD in software research.}
  \label{tab:swot_analysis}
\end{table}

\subsection{Bounded Context and Decomposition of Microservices}
Bounded Context is another fundamental concept in DDD, and its significance in microservices architecture is evident in several studies \cite{maddodi_aggregate_2020, peng_ianticorruption_2007, joselyne_partitioning_2018, garbajosa_supporting_2018, braun_tackling_2021, krause_microservice_2020, pereira_towards_2022, joselyne_systematic_2021, vural_does_2021, bunder_model-driven_2019}. The concept of Bounded Context is used to define explicit boundaries where specific domain models apply, allowing the system to be decomposed into manageable, cohesive units. Decomposing a system into microservices based on Bounded Contexts enables easier scaling, maintenance, and independent updating of each microservice.

\subsection{Domain Services, Entities, and Value Objects as a Tactical DDD}
Several studies discuss the use of Domain Services, Entities, and Value Objects as part of DDD's tactical patterns \cite{maddodi_aggregate_2020, le_generating_2022, garbajosa_supporting_2018, hippchen_designing_2017, ozkan_refactoring_2023, pereira_towards_2022}. Domain Services are employed to handle application tasks and technical code that is not part of the core business logic, while Entities and Value Objects represent different entities and their relationships in the domain model. These concepts contribute to modeling complex business domains and enabling object-oriented representations of domain concepts.

\subsection{ACL for Maintaining Domain Integrity}
When dealing with different bounded contexts, interacting with external systems or refactoring legacy systems, the risk of corrupting a domain model with concepts and structures from another context is high. The ACL acts as a safeguard, ensuring that each context's domain remains untainted and independent. The concept of the ACL is discussed in some studies \cite{peng_ianticorruption_2007, joselyne_partitioning_2018, landre_architectural_2006, le_domain_2018}. It is utilized in the studies to manage dependencies between different domains, preventing the contamination of contexts, safeguard the domain model of a Bounded Context from changes in other contexts, segregating application-specific elements from domain-specific components in microservices and maintaining consistency between aggregates. Studies also suggests that ACL shields new contexts from potential contaminations or disruptions originating in legacy contexts.

\subsection{All Hands on Deck: It is not Only About Software Engineers}
Based on the data extracted from the literature, the implementation of DDD involves the active participation of various key stakeholders who play critical roles in ensuring its successful application. These stakeholders encompass a diverse set of roles, including Sofware Engineers, Architects, Project Managers, and Domain Experts. Software Engineers hold a central position in the DDD implementation process, responsible for translating the domain model into executable code, implementing domain-specific logic, and aligning the software architecture with domain requirements \cite{singjai_patterns_2021, maddodi_aggregate_2020, peng_ianticorruption_2007, snoeck_agile_2022, ozkan_refactoring_2023}. Architects contribute strategic guidance and oversight to the overall software design, collaborating with Software Engineers and other stakeholders to define high-level architecture, bounded contexts, and domain models that effectively capture domain knowledge \cite{maddodi_aggregate_2020, hammoudi_domain-driven_2021, da_silva_bpm2ddd_2022}.

Project Managers play a vital role in coordinating DDD implementation efforts, ensuring alignment with project goals, resource allocation, and effective communication between teams and stakeholders \cite{peng_ianticorruption_2007, garbajosa_supporting_2018}. Domain Experts, possessing in-depth knowledge of the specific business domain, actively contribute to refining the domain model, identifying key domain concepts, and validating the software's alignment with real-world domain requirements \cite{singjai_patterns_2021, garbajosa_supporting_2018, da_silva_bpm2ddd_2022}.

Moreover, the literature also identifies other relevant participants involved in the DDD implementation, including QA Engineers, DevOps Engineers, UX Designers, Business Analysts, Systems Analysts, Data Modeling Specialists, and end-users \cite{peng_ianticorruption_2007, braun_advanced_2021, ozkan_refactoring_2023, oukes_domain-driven_2021, haser_is_2016, bunder_model-driven_2019}. The successful implementation of DDD requires close collaboration and communication among these diverse stakeholders to ensure that the developed software system effectively represents the domain's complexity and meets the business needs. The active engagement of software engineers, architects, project managers, and domain experts, along with the contributions from other relevant participants, enables the application of DDD principles to create robust and domain-centric software solutions. The involvement of these stakeholders enhances the alignment of software systems with real-world domain requirements and fosters the development of high-quality and effective software solutions.

Our review adds a nuanced view of how this collaboration evolves in practice. Rather than treating the developer–domain expert relationship as static, the primary studies reflect varied degrees of maturity in this interaction. For instance, some report structured practices such as Event Storming workshops \cite{garbajosa_supporting_2018}, while others highlight onboarding challenges and communication gaps \cite{ozkan_refactoring_2023, wanderley_framework_2012}. This range of experiences suggests that bridging the gap is not a one-time activity, but an ongoing and often iterative process influenced by tooling, team experience, and organizational context — a perspective less emphasized in prior discussions.

\subsection{Evaluation of DDD Effectiveness}
The evaluation of DDD effectiveness varies across studies. Some studies present empirical evaluations based on metrics, user feedback, or controlled experiments \cite{singjai_patterns_2021, maddodi_aggregate_2020, garbajosa_supporting_2018, braun_tackling_2021, skersys_domain_2012, ozkan_refactoring_2023, mayer_approach_2018}. However, several studies lack specific evaluation results or empirical evidence \cite{perillo_daileon_2009, krause_microservice_2020, ding_enterprise_2020, wanderley_framework_2012, abeck_context_2019, bunder_model-driven_2019}. 

To establish a more comprehensive understanding of the benefits and potential challenges of implementing DDD in different software development scenarios, further empirical research is needed. Researchers and practitioners should focus on collecting concrete metrics, conducting real-world case studies, and measuring the perception of practitioners. We believe that it is also essential to involve human metrics, such as user feedback and assessments from domain experts, architects, and experienced developers. Considering the challenges in implementing DDD, human-centric evaluations can provide valuable insights into the effectiveness of DDD and its impact on the software development process.

\subsection{Rise of the Microservices in DDD}
Our SLR reveals a significant shift in the popularity of microservices in DDD studies starting 2017 and at peak in 2021. As depicted in Figure \ref{fig:publication_year_vs_software_systems}, prior to 2017, credible DDD studies specifically focusing on microservices were scarce, indicating a lack of attention in this area. However, from 2017 onwards, we observed a notable increase in the number of studies exploring the application of DDD principles in the context of microservices. This trend aligns with the findings of a Systematic Mapping Study on Microservices conducted by Hamzehloui et al. \cite{hamzehloui_systematic_2018} in 2018, which also reports a rising number of evaluation papers, suggesting that microservices have been gaining wider implementation in real-world software development scenarios.

Furthermore, Viggiato et al. \cite{viggiato_microservices_2018} in 2018 highlights the increasing popularity of microservices architectures, further emphasizing the growing relevance and adoption of microservices in the software industry. In a similar vein, Knoche et al. \cite{knoche_drivers_2019} in 2019 notes that many companies are actively considering whether microservices are a viable option for their applications, indicating the considerable interest and attention surrounding the adoption of microservices-based architectures.

The convergence of our findings with those of the previously mentioned studies and experts further corroborates the growing significance of microservices in the domain of software development, especially in the context of DDD methodologies. The upward trend in microservices adoption is evident from the year 2017 onwards, indicating a turning point in the landscape of software development practices. Beyond identifying the rise in microservices, our study offers novel insights into how DDD contributes to shaping modularity in such architectures. In particular, we observe that practitioners increasingly rely on DDD patterns like Bounded Contexts and Aggregates to define service boundaries and promote semantic clarity. While prior studies may suggest that microservices benefit from domain-aligned decomposition, our findings ground this assumption in a broader empirical base across 36 peer-reviewed studies. This synthesis provides stronger evidence for how DDD actively influences microservice design practices in both academic and industrial settings.

\subsection{Challenges and Considerations in Implementing DDD Effectively}
To ensure the successful implementation of DDD, it is essential to employ DDD patterns accurately and efficiently. Proper utilization of DDD principles demands a skilled team consisting of experienced developers, domain experts, and seasoned architects \cite{ozkan_refactoring_2023}. The significance of experience in DDD cannot be overstated, as we have observed from various studies that onboarding to DDD is not a straightforward process, and the expertise of developers significantly influences the success of the implementation \cite{hammoudi_domain-driven_2021, ozkan_refactoring_2023}.

Implementing DDD effectively requires a substantial amount of attention and effort, particularly when integrating it into existing software systems. The transition from traditional software development approaches to DDD can be challenging, and the process demands careful planning, training, and a clear understanding of the DDD concepts and principles. Seasoned developers with prior experience in DDD are more likely to grasp the complexities and intricacies of the approach, resulting in better-designed and more maintainable software systems \cite{hippchen_designing_2017}. On the other hand, inexperienced developers may struggle to fully utilize the benefits of DDD, leading to potential issues in the implementation process.

Furthermore, the involvement of domain experts plays a crucial role in the success of DDD implementation. Their deep understanding of the business domain is invaluable in guiding the development team to create accurate and effective domain models \cite{vural_does_2021}. However, acquiring experienced domain experts can be challenging, especially in specialized domains. Their scarcity may pose hurdles in the adoption of DDD, but their involvement is crucial for crafting domain models that truly reflect the intricacies and requirements of the real-world domain.

\subsection{Academia-Industry Collaboration}
The academic studies in our SLR predominantly emphasize the theoretical aspects of DDD. They explore the foundational principles and patterns of DDD, such as Ubiquitous Language, Bounded Context, Entities, and Value Objects. These studies often propose novel frameworks, tools and methodologies that extend the DDD concepts to specific problem domains. While they provide valuable insights into the theoretical underpinnings of DDD, they might lack empirical validation and practical implementation in real-world projects.

On the other hand, we experienced that the industry-focused studies put DDD principles into practice in various software systems. They showcase how DDD is applied to real-world projects and explore its impact on system design, architecture, and development. These studies often involve case studies and experience reports from software engineers, architects, and domain experts. The emphasis is on practical implementation, challenges faced, and lessons learned while applying DDD in complex software projects.

Our findings indicate a healthy mix of academic and industry studies, reflecting the growing interest in DDD from both research and practical application perspectives. While academic studies contribute theoretical advancements and innovative approaches, industry-focused studies demonstrate the real-world applicability and effectiveness of DDD principles in solving complex software design and development challenges. More interaction between academia and industry in DDD research could potentially facilitate the transfer of knowledge and best practices, contributing to the further advancement and adoption of DDD principles in real-world software development scenarios, as successfully demonstrated in \cite{ozkan_refactoring_2023}.

\subsubsection {Contrasting Academic and Industry Perspectives}
While our study includes a balanced set of academic (58.3\%) and industry (36.1\%) studies—with a small number of hybrid cases (5.6\%)—the practical perspectives from industry deserve further distinction. Academic research on DDD tends to focus on conceptual modeling, tool development (e.g., DSLs, context mapping tools), and theoretical discussions on patterns such as Bounded Contexts or Entities. These studies often propose frameworks or methodologies but may lack concrete empirical validation. In contrast, industry studies are generally more experience-driven, describing real-world applications of DDD in large-scale systems such as microservices, enterprise refactoring, or system modernization. These studies prioritize architectural pragmatism, performance concerns, and integration challenges, and often highlight the social or organizational implications of DDD adoption (e.g., communication, onboarding, and team alignment).

For example, the academic studies often examine how to formalize context boundaries using modeling tools like Context Mapper or domain-specific languages, whereas the industry studies report practical experience applying bounded contexts in cross-functional teams or resolving communication issues between developers and domain experts. Moreover, industry papers tend to use anecdotal or case-based evidence, while academic studies more frequently advocate for abstract, generalizable models. This contrast reveals complementary strengths: academia offers methodological structure and design guidance, while industry grounds these concepts in complex, evolving development environments. Recognizing this distinction helps to contextualize the relevance and maturity of DDD research across settings.

\subsection{A Critical Analysis of Research Quality in the DDD Literature}
The analyzed studies demonstrate varying degrees of adherence to DDD principles and the effective application of DDD concepts. While some studies effectively apply DDD principles, others exhibit shortcomings in implementation and evaluation. To enhance research quality, future studies should emphasize the consistent use of Ubiquitous Language, explicit definition of Bounded Contexts, proper utilization of Domain Events and DDD patterns, and robust evaluation methodologies. Additionally, studies should openly discuss the challenges and limitations of DDD, contributing to a more comprehensive understanding of its potential and practical implications in software development.

Several studies employed the notion of Ubiquitous Language, which facilitates effective communication between stakeholders and software designers by creating a shared vocabulary. However, a few studies failed to implement this language consistently, potentially hindering communication and understanding among team members. Future research should emphasize the importance of adopting and consistently applying Ubiquitous Language throughout the development process to ensure better alignment and clarity.

Bounded Context, another key concept in DDD, helps define explicit boundaries where specific domain models apply \cite{evans_domain-driven_2004, vernon_domain-driven_2016}. While some studies effectively utilized Bounded Context to organize and segregate domains, others did not explicitly mention or implement this principle. Ensuring clear identification and definition of Bounded Contexts is crucial for managing interactions between different models and contexts, and future studies should focus on this aspect to enhance research quality.

The application of Domain Events to trigger behavior across multiple aggregates within a Bounded Context was observed in some studies. However, not all studies effectively leveraged this concept. Implementing Domain Events can improve event-driven communication and coordination between domain components, promoting a more cohesive and flexible architecture \cite{evans_domain-driven_2004, vernon_domain-driven_2016}. Future research should explore the potential benefits of Domain Events in event-driven systems and provide concrete examples of their implementation.

Several studies showcased the use of Aggregates, Entities, and Value Objects to model complex business domains. While these concepts were well-implemented in some cases, others lacked specific details or provided inadequate explanations about their usage. To bolster research quality, studies should strive for comprehensive and consistent implementation of these DDD patterns, ensuring a clear understanding of their roles and relationships within the domain model.

We observed that Strategic Design patterns such as Core Domain, Shared Kernel, and upstream–downstream relationships were underreported in most studies. This suggests a possible gap in the literature, where tactical DDD patterns are emphasized, but strategic alignment is less frequently explored or documented. Future research should pay more attention to these higher-level design patterns to understand their role in system architecture and team collaboration.

One significant aspect that emerged from the data was the lack of specific evaluation methods in some studies. While some research incorporated empirical evaluations, others relied solely on proposing methodologies or demonstrating case studies without conducting rigorous evaluations. Future studies should adopt more robust evaluation methods, such as controlled experiments, surveys, or user feedback, to assess the effectiveness of DDD implementation and provide empirical evidence to support their findings.

Moreover, most studies did not discuss the challenges and limitations associated with DDD adoption, leaving potential gaps in addressing concerns or drawbacks of the approach. To improve research quality, studies should acknowledge and discuss both the advantages and limitations of DDD, offering a balanced view to readers and practitioners. This will help in understanding the practical implications of DDD and guide future research and implementation efforts.

Future research can provide a more comprehensive and balanced analysis of DDD by addressing these areas to contribute to the advancement of knowledge and practice in the field of software development.

\section{Implications}
In this section, we introduce the implications of our findings for both practitioners and researchers. We draw on the results presented in our RQs to ground our implications in the evidence synthesized from the primary studies.

\textbf{Implications for Practitioners.}  
One of the primary implications for practitioners is the recognition of the diverse benefits that DDD can bring to various software systems. The application of DDD principles has been shown to improve the modularity, maintainability, and scalability of microservices-based applications (Section \ref{subsec:rq3}), thereby enabling organizations to handle complex and distributed systems more effectively~\cite{singjai_patterns_2021, joselyne_partitioning_2018, wang_reference_2022}. Practitioners can leverage these insights to enhance their software architecture and development practices by adopting DDD patterns such as Bounded Contexts and Aggregates to define clear service boundaries and promote decoupling.

Moreover, the adoption of DDD in enterprise software systems has proven valuable in addressing challenges related to legacy system modernization and increasing architectural complexity \cite{maddodi_aggregate_2020, peng_ianticorruption_2007, garbajosa_supporting_2018}. Section \ref{subsec:rq2} illustrates how DDD facilitates flexible and maintainable architectures by aligning software components with evolving business needs. Practitioners aiming to modernize legacy systems should consider using context maps and ACLs (as highlighted in Section \ref{subsec:rq5}) to incrementally isolate and refactor legacy code while preserving domain integrity.

Additionally, DDD has demonstrated benefits in improving code quality and software construction in web applications and development tools \cite{le_generating_2022, perillo_daileon_2009, kapferer_domain-driven_2020}. In Section \ref{subsec:rq3}, studies report that domain annotations and modeling tools help practitioners maintain separation of concerns and automate parts of the development process. These insights can guide practitioners in selecting DDD-aligned tools that improve code maintainability \cite{le_generating_2022} and streamline software generation.

However, DDD is not without its challenges. As discussed in Section \ref{subsec:rq5}, one of the most significant obstacles is the steep learning curve, especially for developers new to DDD~\cite{ozkan_refactoring_2023, hippchen_designing_2017}. Misunderstandings in the application of patterns, unclear decomposition strategies, and the misalignment between teams and domains are common pitfalls. To mitigate this, practitioners should invest in onboarding initiatives, including dedicated DDD training sessions, domain modeling workshops (e.g., Event Storming \cite{garbajosa_supporting_2018}), and mentorship from experienced architects (see Table \ref{tab:ddd_challenges_mitigations}).

\textbf{Implications for Researchers.}  
Despite the progress made in understanding the application and benefits of DDD, Section \ref{subsec:how_effectiveness_measured} shows that only a limited number of studies (approximately 39\%) employ empirical evaluations. Future research should prioritize the design of empirical studies with concrete evaluation metrics—such as team productivity, domain alignment, and architectural complexity—to measure the actual impact of DDD adoption in industrial settings~\cite{singjai_patterns_2021, maddodi_aggregate_2020, garbajosa_supporting_2018}.

Furthermore, Section \ref{subsec:rq5} indicates that the complexity of managing domain boundaries, integrating DDD with legacy systems, and supporting DSL-based tooling are persistent technical challenges. Researchers should explore novel techniques—e.g., automated decomposition, model-code synchronization, and refactoring strategies—to support scalable DDD adoption in complex environments \cite{braun_tackling_2021, ozkan_refactoring_2023}.

Another important research direction involves studying stakeholder dynamics. Section \ref{subsec:rq7} emphasizes the critical role of collaboration between software engineers, architects, and domain experts in the success of DDD. Yet, communication breakdowns and a lack of domain understanding remain major barriers \cite{wanderley_framework_2012, da_silva_bpm2ddd_2022}. Researchers could investigate socio-technical methods (e.g., participatory modeling \cite{voinov_modeling_2010, brandolini_eventstorming_2019}, collaborative tooling \cite{famelis_partial_2012}) to bridge this gap and formalize co-design practices.

In addition, Section \ref{subsec:how_effectiveness_measured} reveals that studies vary widely in how they evaluate DDD. Developing standardized frameworks and reproducible evaluation methodologies will help unify the field and allow comparison across contexts \cite{ozkan_refactoring_2023, le_generating_2022}. Incorporating both technical metrics (e.g., coupling, cohesion) and human-centered metrics (e.g., developer perception, onboarding effort) can lead to a more holistic understanding of DDD's effectiveness.

Based on our findings, future research should prioritize empirical studies that not only report qualitative insights but also collect large-scale (e.g.~via repository mining for DDD similar to~\cite{babur2024language}), longitudinal and team-level data. While several studies use focus groups and interviews, we observed a lack of sustained evaluations on aspects such as maintainability, modularity drift over time, or onboarding effectiveness. Studies like Braun et. al. \cite{braun_tackling_2021,braun_advanced_2021} and \"{O}zkan et. al. \cite{ozkan_refactoring_2023} demonstrate the potential of combining qualitative feedback with structured metrics, yet these efforts remain isolated. To build on this, researchers could design action research or longitudinal case studies that track DDD adoption over multiple sprints or releases, focusing on how domain alignment evolves, how design debt is managed, and how development teams adapt. Similarly, design-science-based evaluations of modeling tools (e.g., Context Mapper) can shed light on how tooling impacts the domain-code connection in practical workflows. These directions would help close both methodological and tooling-related gaps identified in current literature.

By addressing these research gaps, scholars can contribute not only to the theoretical advancement of DDD but also to practical guidelines, tool support, and empirical validation frameworks that help teams successfully adopt DDD in complex software projects.

\section{Limitations and Threats to Validity}
Some limitations and validity considerations are applicable for SLR studies \cite{petersen_guidelines_2015, petersen_systematic_2008}. The limitations and threats to the validity of this study are discussed in this section.

\textbf{Timeframe and Currency.} One limitation is the timeframe of the literature search and the currency of the included studies. The SLR is based on data available up to June 2023, and relevant studies published after that date might have been missed. The field of DDD is rapidly evolving, and newer developments or trends may not be fully captured in this review.

\textbf{Quality of Studies.} The quality of individual studies can impact the overall reliability of the SLR findings. The included studies vary in terms of methodological rigor, sample size, and research design. Some studies might have limitations or biases that could influence their results and subsequent conclusions. Nonetheless, we endeavored to rigorously exclude studies of low quality by systematically applying inclusion and exclusion criteria as well as conducting thorough quality assessments.

\textbf{Inclusion and Exclusion Criteria.}
The effectiveness of an SLR largely depends on the clarity and appropriateness of the inclusion and exclusion criteria. Despite our efforts to define comprehensive criteria, there is a possibility of unintentionally excluding relevant studies or including studies that do not precisely fit the research questions. To mitigate this potential limitation, a rigorous approach was adopted. Two authors independently applied predefined inclusion and exclusion criteria to a randomly selected pool of 60 studies. The level of agreement between the two authors was assessed using Cohen's Kappa coefficient \cite{cohen_coefficient_1960}, a widely accepted measure of inter-rater reliability. Encouragingly, the analysis revealed a perfect agreement between the authors in their decisions regarding study inclusion and exclusion within the randomly sampled studies. This robust agreement underscores the reliability of the selection process and enhances the credibility of the findings.

\textbf{Bias in Data Extraction.} A limitation related to the validity of data extraction in this study is the potential for human error or bias during the data extraction process. Despite efforts to maintain consistency and accuracy, the interpretation of information from primary studies could be influenced by the subjectivity of the researchers. Different individuals extracting data may interpret the findings differently or unintentionally introduce errors, leading to inconsistencies in the extracted data. To address this limitation, a careful and systematic approach was employed, and multiple researchers were involved in the data extraction process. Inter-rater reliability checks and discussions among the researchers were conducted to ensure agreement and minimize discrepancies. However, it is essential to acknowledge that the possibility of human-related errors cannot be entirely eliminated, and they may have some impact on the final results and conclusions of the systematic literature review.

\textbf{Publication Bias.} One potential threat to validity is publication bias. The studies included in this SLR were obtained from various academic databases and sources, which might not represent the entire body of literature on DDD. Some studies might not have been published or accessible, leading to a skewed representation of the research landscape. Nevertheless, to mitigate this bias, we performed a comprehensive search across multiple academic databases, including Google Scholar and Scopus during the snowballing process. This extensive search approach was aimed at encompassing a wide range of academic sources and reducing potential biases in the selection of studies.

\textbf{Language Bias.} The SLR focused primarily on studies published in English, which might introduce a language bias. Relevant research conducted in other languages could have been omitted, potentially impacting the comprehensiveness of the review.

\textbf{Agreement Level During Quality Assessment.} Another potential threat to validity is the lack of assessment of agreement levels between assessors during the quality assessment stage. While Cohen's Kappa was used to evaluate inter-rater reliability during the inclusion/exclusion stage, no similar measure was applied during the quality assessment of the included studies. This omission could introduce subjectivity and potential bias in the quality assessment process, as individual assessors may have differing interpretations of quality criteria. However, during the quality assessment stage, in cases of disagreement or doubt, the assessors consulted with each other to reach a consensus. This approach used to ensure that the quality assessments were rigorous, minimized potential biases, and allowed discrepancies to be resolved systematically.

\textbf{Study-Specific Validity Considerations.}  
In addition to general methodological considerations, we acknowledge several threats specific to our study. First, our focus on implementation-related search terms (e.g., \textit{refactor*, improv*, optimiz*}) may have led to the exclusion of relevant DDD papers that are more theoretical or descriptive in nature. This trade-off was a deliberate decision to ensure practical relevance, but it may affect the completeness of our study population. Second, while our inclusion of forward snowballing from Evans' seminal book aimed to enrich the dataset with practically grounded work, it may also have introduced a bias toward studies that cite this particular source, possibly underrepresenting alternative DDD schools of thought. Finally, our coding and classification of DDD patterns — especially in distinguishing strategic vs. tactical applications — required interpretation and contextual judgment. Despite using a predefined form and multiple authors reviewing the coding, subjectivity may still have influenced the final categorization.

\section{Conclusion}
This SLR provides a comprehensive overview of the effectiveness, challenges, and implications of adopting DDD in software development. Through the analysis of a diverse set of primary studies, we identified key stakeholders involved in the implementation of DDD, including Software Engineers, Architects, Project Managers, and Domain Experts. These stakeholders play critical roles in ensuring the successful application of DDD principles and practices in various software systems.

The findings highlight the significance of Ubiquitous Language as a core principle of DDD, emphasizing its role in facilitating effective communication and collaboration among team members. Bounded Contexts and the decomposition of microservices stand out as fundamental concepts in DDD, enabling the creation of scalable and maintainable software systems. Domain Services, Entities, and Value Objects serve as tactical DDD patterns, contributing to the modeling of complex business domains.

The implementation of DDD demands considerable attention and effort, particularly in transitioning from traditional software development approaches to DDD. Experienced developers and domain experts play crucial roles in effectively implementing DDD concepts. The involvement of domain experts is especially critical for crafting domain models that accurately represent real-world domain requirements.

The SLR also highlights the rise of microservices in the context of DDD, indicating a significant shift in the popularity of microservices-based architectures since 2017. However, further empirical research is needed to comprehensively understand the benefits and challenges of implementing DDD in different software development scenarios.

The collaboration between academia and industry in DDD research is evident, with academic studies contributing theoretical advancements and innovative approaches, while industry-focused studies showcase practical implementation and real-world applicability. This collaboration can facilitate knowledge transfer and best practices, contributing to the advancement and adoption of DDD principles in software development.

A critical analysis of research quality in the DDD literature reveals varying degrees of adherence to DDD principles and the effective application of DDD concepts. Future studies should emphasize the consistent use of Ubiquitous Language, clear definition of Bounded Contexts, proper utilization of Domain Events and DDD patterns, and robust evaluation methodologies. Moreover, studies should openly discuss the challenges and limitations of DDD to provide a comprehensive understanding of its potential and practical implications.

DDD offers valuable insights and practices for software development, fostering communication, and collaboration among stakeholders while modeling complex business domains. However, its successful implementation depends on the expertise of stakeholders, proper utilization of DDD principles, and the adoption of robust evaluation methods. As software systems continue to grow in complexity, DDD remains a relevant and promising approach for developing domain-centric and scalable software solutions.

\section*{Declaration of Competing Interest}
The authors declare that they have no known competing financial interests or personal relationships that could have appeared to influence the work reported in this paper.

\section*{Data Availability}
Data will be made available on request.

\bibliographystyle{elsarticle-num-names}
\bibliography{references}

\begin{thebibliography}{71}
\expandafter\ifx\csname natexlab\endcsname\relax\def\natexlab#1{#1}\fi
\providecommand{\url}[1]{\texttt{#1}}
\providecommand{\href}[2]{#2}
\providecommand{\path}[1]{#1}
\providecommand{\DOIprefix}{doi:}
\providecommand{\ArXivprefix}{arXiv:}
\providecommand{\URLprefix}{URL: }
\providecommand{\Pubmedprefix}{pmid:}
\providecommand{\doi}[1]{\href{http://dx.doi.org/#1}{\path{#1}}}
\providecommand{\Pubmed}[1]{\href{pmid:#1}{\path{#1}}}
\providecommand{\bibinfo}[2]{#2}
\ifx\xfnm\relax \def\xfnm[#1]{\unskip,\space#1}\fi
\bibitem[{Avram and Marinescu(2006)}]{avram_domain-driven_2006}
\bibinfo{author}{A.~Avram}, \bibinfo{author}{F.~Marinescu},
  \bibinfo{title}{Domain-{{Driven Design Quickly}}: {{A Summary}} of {{Eric
  Evans}}' {{Domain-Driven Design}}}, Enterprise {{Software Development
  Series}}, \bibinfo{address}{C4Media}, \bibinfo{year}{2006}.
\bibitem[{Evans(2004)}]{evans_domain-driven_2004}
\bibinfo{author}{E.~Evans}, \bibinfo{title}{Domain-Driven Design: Tackling
  Complexity in the Heart of Software}, \bibinfo{publisher}{Addison-Wesley},
  \bibinfo{address}{Boston}, \bibinfo{year}{2004}.
\bibitem[{Vernon(2013)}]{vernon_implementing_2013}
\bibinfo{author}{V.~Vernon}, \bibinfo{title}{Implementing Domain-Driven
  Design}, \bibinfo{publisher}{Addision-Wesley}, \bibinfo{address}{Upper Saddle
  River, NJ}, \bibinfo{year}{2013}.
\bibitem[{Kitchenham and Charters(2007)}]{kitchenham_guidelines_2007}
\bibinfo{author}{B.~Kitchenham}, \bibinfo{author}{S.~Charters},
\newblock \bibinfo{title}{Guidelines for performing {{Systematic Literature
  Reviews}} in {{Software Engineering}}} \bibinfo{volume}{2}
  (\bibinfo{year}{2007}).
\bibitem[{Wohlin(2014)}]{wohlin_guidelines_2014}
\bibinfo{author}{C.~Wohlin},
\newblock \bibinfo{title}{Guidelines for snowballing in systematic literature
  studies and a replication in software engineering},
\newblock in: \bibinfo{booktitle}{Proceedings of the 18th {{International
  Conference}} on {{Evaluation}} and {{Assessment}} in {{Software
  Engineering}}}, \bibinfo{publisher}{ACM}, \bibinfo{address}{London England
  United Kingdom}, \bibinfo{year}{2014}, pp. \bibinfo{pages}{1--10}.
  \DOIprefix\doi{10.1145/2601248.2601268}.
\bibitem[{Gamma et~al.(2009)Gamma, Helm, Johnson, and
  Vlissides}]{gamma_design_2009}
\bibinfo{author}{E.~Gamma}, \bibinfo{author}{R.~Helm}, \bibinfo{author}{R.~E.
  Johnson}, \bibinfo{author}{J.~Vlissides}, \bibinfo{title}{Design patterns:
  elements of reusable object-oriented software},
  \bibinfo{publisher}{Addison-Wesley}, \bibinfo{address}{Reading, MA},
  \bibinfo{year}{2009}. \bibinfo{note}{OCLC: 624525693}.
\bibitem[{Buschmann(1996)}]{buschmann_pattern-oriented_1996}
\bibinfo{editor}{F.~Buschmann} (Ed.), \bibinfo{title}{Pattern-oriented software
  architecture: a system of patterns}, \bibinfo{publisher}{Wiley},
  \bibinfo{address}{Chichester ; New York}, \bibinfo{year}{1996}.
\bibitem[{Brandolini(2013)}]{brandolini_eventstorming_2013}
\bibinfo{author}{A.~Brandolini}, \bibinfo{title}{Introducing eventstorming –
  an explosive modeling technique for lean discovery},
  \bibinfo{howpublished}{\url{https://ziobrando.blogspot.com/2013/11/introducing-eventstorming.html}},
  \bibinfo{year}{2013}. \bibinfo{note}{Accessed April 2025}.
\bibitem[{Le et~al.(2016)Le, Dang, and Nguyen}]{le_domain-driven_2016}
\bibinfo{author}{D.~M. Le}, \bibinfo{author}{D.-H. Dang},
  \bibinfo{author}{V.-H. Nguyen},
\newblock \bibinfo{title}{Domain-driven design patterns: {{A}} metadata-based
  approach},
\newblock in: \bibinfo{booktitle}{2016 {{IEEE RIVF International Conference}}
  on {{Computing}} \& {{Communication Technologies}}, {{Research}},
  {{Innovation}}, and {{Vision}} for the {{Future}} ({{RIVF}})},
  \bibinfo{publisher}{IEEE}, \bibinfo{address}{Hanoi}, \bibinfo{year}{2016},
  pp. \bibinfo{pages}{247--252}. \DOIprefix\doi{10.1109/RIVF.2016.7800302}.
\bibitem[{Singjai et~al.(2021)Singjai, Zdun, and
  Zimmermann}]{singjai_practitioner_2021}
\bibinfo{author}{A.~Singjai}, \bibinfo{author}{U.~Zdun},
  \bibinfo{author}{O.~Zimmermann},
\newblock \bibinfo{title}{Practitioner {{Views}} on the {{Interrelation}} of
  {{Microservice APIs}} and {{Domain-Driven Design}}: {{A Grey Literature Study
  Based}} on {{Grounded Theory}}},
\newblock in: \bibinfo{booktitle}{2021 {{IEEE}} 18th {{International
  Conference}} on {{Software Architecture}} ({{ICSA}})},
  \bibinfo{publisher}{IEEE}, \bibinfo{address}{Stuttgart, Germany},
  \bibinfo{year}{2021}, pp. \bibinfo{pages}{25--35}.
  \DOIprefix\doi{10.1109/ICSA51549.2021.00011}.
\bibitem[{De~Goey et~al.(2019)De~Goey, Hilletofth, and
  Eriksson}]{de_goey_design-driven_2019}
\bibinfo{author}{H.~De~Goey}, \bibinfo{author}{P.~Hilletofth},
  \bibinfo{author}{L.~Eriksson},
\newblock \bibinfo{title}{Design-driven innovation: A systematic literature
  review},
\newblock \bibinfo{journal}{European Business Review} \bibinfo{volume}{31}
  (\bibinfo{year}{2019}) \bibinfo{pages}{92--114}.
  \DOIprefix\doi{10.1108/EBR-09-2017-0160}.
\bibitem[{Bertoni(2020)}]{bertoni_data-driven_2020}
\bibinfo{author}{A.~Bertoni},
\newblock \bibinfo{title}{{{DATA-DRIVEN DESIGN IN CONCEPT DEVELOPMENT}}:
  {{SYSTEMATIC REVIEW AND MISSED OPPORTUNITIES}}},
\newblock \bibinfo{journal}{Proceedings of the Design Society: DESIGN
  Conference} \bibinfo{volume}{1} (\bibinfo{year}{2020})
  \bibinfo{pages}{101--110}. \DOIprefix\doi{10.1017/dsd.2020.4}.
\bibitem[{Stol and Fitzgerald(2018)}]{stol2018abc}
\bibinfo{author}{K.-J. Stol}, \bibinfo{author}{B.~Fitzgerald},
\newblock \bibinfo{title}{The {ABC} of software engineering research},
\newblock \bibinfo{journal}{ACM Transactions on Software Engineering and
  Methodology (TOSEM)} \bibinfo{volume}{27} (\bibinfo{year}{2018})
  \bibinfo{pages}{1--51}.
\bibitem[{L{\"u}bke et~al.(2019)L{\"u}bke, Zimmermann, Pautasso, Zdun, and
  Stocker}]{lubke_interface_2019}
\bibinfo{author}{D.~L{\"u}bke}, \bibinfo{author}{O.~Zimmermann},
  \bibinfo{author}{C.~Pautasso}, \bibinfo{author}{U.~Zdun},
  \bibinfo{author}{M.~Stocker},
\newblock \bibinfo{title}{Interface {{Evolution Patterns}}: {{Balancing
  Compatibility}} and {{Extensibility}} across {{Service Life Cycles}}},
\newblock in: \bibinfo{booktitle}{Proceedings of the 24th {{European
  Conference}} on {{Pattern Languages}} of {{Programs}}}, {{EuroPLop}} '19,
  \bibinfo{publisher}{Association for Computing Machinery},
  \bibinfo{address}{New York, NY, USA}, \bibinfo{year}{2019}.
  \DOIprefix\doi{10.1145/3361149.3361164}.
\bibitem[{Giray et~al.(2023)Giray, Bennin, K{\"o}ksal, Babur, and
  Tekinerdogan}]{giray_use_2023}
\bibinfo{author}{G.~Giray}, \bibinfo{author}{K.~E. Bennin},
  \bibinfo{author}{{\"O}.~K{\"o}ksal}, \bibinfo{author}{{\"O}.~Babur},
  \bibinfo{author}{B.~Tekinerdogan},
\newblock \bibinfo{title}{On the use of deep learning in software defect
  prediction},
\newblock \bibinfo{journal}{Journal of Systems and Software}
  \bibinfo{volume}{195} (\bibinfo{year}{2023}) \bibinfo{pages}{111537}.
  \URLprefix
  \url{https://linkinghub.elsevier.com/retrieve/pii/S0164121222002138}.
  \DOIprefix\doi{10.1016/j.jss.2022.111537}.
\bibitem[{Ochoa et~al.(2025)Ochoa, Hammad, Giray, Babur, and
  Bennin}]{ochoa2025characterising}
\bibinfo{author}{L.~Ochoa}, \bibinfo{author}{M.~Hammad},
  \bibinfo{author}{G.~Giray}, \bibinfo{author}{{\"O}.~Babur},
  \bibinfo{author}{K.~Bennin},
\newblock \bibinfo{title}{Characterising harmful api uses and repair
  techniques: Insights from a systematic review},
\newblock \bibinfo{journal}{Computer Science Review} \bibinfo{volume}{57}
  (\bibinfo{year}{2025}) \bibinfo{pages}{100732}.
\bibitem[{Kuhrmann et~al.(2017)Kuhrmann, Fern{\'a}ndez, and
  Daneva}]{kuhrmann_pragmatic_2017}
\bibinfo{author}{M.~Kuhrmann}, \bibinfo{author}{D.~M. Fern{\'a}ndez},
  \bibinfo{author}{M.~Daneva},
\newblock \bibinfo{title}{On the pragmatic design of literature studies in
  software engineering: An experience-based guideline},
\newblock \bibinfo{journal}{Empirical Software Engineering}
  \bibinfo{volume}{22} (\bibinfo{year}{2017}) \bibinfo{pages}{2852--2891}.
  \DOIprefix\doi{10.1007/s10664-016-9492-y}.
\bibitem[{Cohen(1960)}]{cohen_coefficient_1960}
\bibinfo{author}{J.~Cohen},
\newblock \bibinfo{title}{A {{Coefficient}} of {{Agreement}} for {{Nominal
  Scales}}},
\newblock \bibinfo{journal}{Educational and Psychological Measurement}
  \bibinfo{volume}{20} (\bibinfo{year}{1960}) \bibinfo{pages}{37--46}.
  \DOIprefix\doi{10.1177/001316446002000104}.
\bibitem[{Torres et~al.(2021)Torres, Van~Den~Brand, and
  Serebrenik}]{torres_systematic_2021}
\bibinfo{author}{W.~Torres}, \bibinfo{author}{M.~G.~J. Van~Den~Brand},
  \bibinfo{author}{A.~Serebrenik},
\newblock \bibinfo{title}{A systematic literature review of cross-domain model
  consistency checking by model management tools},
\newblock \bibinfo{journal}{Software and Systems Modeling} \bibinfo{volume}{20}
  (\bibinfo{year}{2021}) \bibinfo{pages}{897--916}.
  \DOIprefix\doi{10.1007/s10270-020-00834-1}.
\bibitem[{Wang et~al.(2023)Wang, Huang, Gao, Ge, Zhang, Feng, Satyarth, Li,
  Zhang, and Ng}]{wang_machinedeep_2023}
\bibinfo{author}{S.~Wang}, \bibinfo{author}{L.~Huang},
  \bibinfo{author}{A.~Gao}, \bibinfo{author}{J.~Ge},
  \bibinfo{author}{T.~Zhang}, \bibinfo{author}{H.~Feng},
  \bibinfo{author}{I.~Satyarth}, \bibinfo{author}{M.~Li},
  \bibinfo{author}{H.~Zhang}, \bibinfo{author}{V.~Ng},
\newblock \bibinfo{title}{Machine/{{Deep Learning}} for {{Software
  Engineering}}: {{A Systematic Literature Review}}},
\newblock \bibinfo{journal}{IEEE Transactions on Software Engineering}
  \bibinfo{volume}{49} (\bibinfo{year}{2023}) \bibinfo{pages}{1188--1231}.
  \DOIprefix\doi{10.1109/TSE.2022.3173346}.
\bibitem[{Vernon(2016)}]{vernon_domain-driven_2016}
\bibinfo{author}{V.~Vernon}, \bibinfo{title}{Domain-Driven Design Distilled},
  \bibinfo{publisher}{Addison-Wesley}, \bibinfo{address}{Boston},
  \bibinfo{year}{2016}.
\bibitem[{Nilsson(2006)}]{nilsson_applying_2006}
\bibinfo{author}{J.~Nilsson}, \bibinfo{title}{Applying Domain-Driven Design and
  Patterns: With Examples in {{C}}\# and .{{NET}}},
  \bibinfo{publisher}{Addison-Wesley}, \bibinfo{address}{Upper Saddle River,
  NJ}, \bibinfo{year}{2006}.
\bibitem[{Popay et~al.(2006)Popay, Roberts, Sowden, Petticrew, Arai, Rodgers,
  Britten, Roen, and Duffy}]{popay_guidance_2006}
\bibinfo{author}{J.~Popay}, \bibinfo{author}{H.~Roberts},
  \bibinfo{author}{A.~Sowden}, \bibinfo{author}{M.~Petticrew},
  \bibinfo{author}{L.~Arai}, \bibinfo{author}{M.~Rodgers},
  \bibinfo{author}{N.~Britten}, \bibinfo{author}{K.~Roen},
  \bibinfo{author}{S.~Duffy}, \bibinfo{title}{Guidance on the Conduct of
  Narrative Synthesis in Systematic Reviews: {{A}} Product from the {{ESRC
  Methods Programme}}}, \bibinfo{publisher}{Lancaster University},
  \bibinfo{year}{2006}. \DOIprefix\doi{10.13140/2.1.1018.4643}.
\bibitem[{Singjai et~al.(2021)Singjai, Zdun, Zimmermann, and
  Pautasso}]{singjai_patterns_2021}
\bibinfo{author}{A.~Singjai}, \bibinfo{author}{U.~Zdun},
  \bibinfo{author}{O.~Zimmermann}, \bibinfo{author}{C.~Pautasso},
\newblock \bibinfo{title}{Patterns on {{Deriving APIs}} and their {{Endpoints}}
  from {{Domain Models}}},
\newblock in: \bibinfo{booktitle}{26th {{European Conference}} on {{Pattern
  Languages}} of {{Programs}}}, \bibinfo{publisher}{ACM},
  \bibinfo{address}{Graz Austria}, \bibinfo{year}{2021}, pp.
  \bibinfo{pages}{1--15}. \DOIprefix\doi{10.1145/3489449.3489976}.
\bibitem[{Maddodi et~al.(2020)Maddodi, Jansen, and
  Overeem}]{maddodi_aggregate_2020}
\bibinfo{author}{G.~Maddodi}, \bibinfo{author}{S.~Jansen},
  \bibinfo{author}{M.~Overeem},
\newblock \bibinfo{title}{Aggregate {{Architecture Simulation}} in
  {{Event-Sourcing Applications}} using {{Layered Queuing Networks}}},
\newblock in: \bibinfo{booktitle}{Proceedings of the {{ACM}}/{{SPEC
  International Conference}} on {{Performance Engineering}}},
  \bibinfo{publisher}{ACM}, \bibinfo{address}{Edmonton AB Canada},
  \bibinfo{year}{2020}, pp. \bibinfo{pages}{238--245}.
  \DOIprefix\doi{10.1145/3358960.3375797}.
\bibitem[{Peng and Hu(2007)}]{peng_ianticorruption_2007}
\bibinfo{author}{S.~Peng}, \bibinfo{author}{Y.~Hu},
\newblock \bibinfo{title}{{{IAnticorruption}}: A domain-driven design approach
  to more robust integration},
\newblock in: \bibinfo{booktitle}{Companion to the 22nd {{ACM SIGPLAN}}
  Conference on {{Object-oriented}} Programming Systems and Applications
  Companion}, \bibinfo{publisher}{ACM}, \bibinfo{address}{Montreal Quebec
  Canada}, \bibinfo{year}{2007}, pp. \bibinfo{pages}{976--982}.
  \DOIprefix\doi{10.1145/1297846.1297966}.
\bibitem[{Jos{\'e}lyne et~al.(2018)Jos{\'e}lyne, {Tuheirwe-Mukasa}, Kanagwa,
  and Balikuddembe}]{joselyne_partitioning_2018}
\bibinfo{author}{M.~I. Jos{\'e}lyne}, \bibinfo{author}{D.~{Tuheirwe-Mukasa}},
  \bibinfo{author}{B.~Kanagwa}, \bibinfo{author}{J.~Balikuddembe},
\newblock \bibinfo{title}{Partitioning microservices: A domain engineering
  approach},
\newblock in: \bibinfo{booktitle}{Proceedings of the 2018 {{International
  Conference}} on {{Software Engineering}} in {{Africa}}},
  \bibinfo{publisher}{ACM}, \bibinfo{address}{Gothenburg Sweden},
  \bibinfo{year}{2018}, pp. \bibinfo{pages}{43--49}.
  \DOIprefix\doi{10.1145/3195528.3195535}.
\bibitem[{Perillo et~al.(2009)Perillo, Guerra, and
  Fernandes}]{perillo_daileon_2009}
\bibinfo{author}{J.~R.~C. Perillo}, \bibinfo{author}{E.~M. Guerra},
  \bibinfo{author}{C.~T. Fernandes},
\newblock \bibinfo{title}{Daileon: A tool for enabling domain annotations},
\newblock in: \bibinfo{booktitle}{Proceedings of the {{Workshop}} on {{AOP}}
  and {{Meta-Data}} for {{Software Evolution}}}, \bibinfo{publisher}{ACM},
  \bibinfo{address}{Genova Italy}, \bibinfo{year}{2009}, pp.
  \bibinfo{pages}{1--4}. \DOIprefix\doi{10.1145/1562860.1562867}.
\bibitem[{Le et~al.(2019)Le, Dang, and Vu}]{le_jdomainapp_2019}
\bibinfo{author}{D.~M. Le}, \bibinfo{author}{D.-H. Dang},
  \bibinfo{author}{H.~T. Vu},
\newblock \bibinfo{title}{{{jDomainApp}}: {{A Module-Based Domain-Driven
  Software Framework}}},
\newblock in: \bibinfo{booktitle}{Proceedings of the {{Tenth International
  Symposium}} on {{Information}} and {{Communication Technology}} - {{SoICT}}
  2019}, \bibinfo{publisher}{ACM Press}, \bibinfo{address}{Hanoi, Ha Long Bay,
  Viet Nam}, \bibinfo{year}{2019}, pp. \bibinfo{pages}{399--406}.
  \DOIprefix\doi{10.1145/3368926.3369657}.
\bibitem[{Le et~al.(2022)Le, Nguyen, Tran, Le, and Tong}]{le_generating_2022}
\bibinfo{author}{D.~M. Le}, \bibinfo{author}{A.~P. Nguyen},
  \bibinfo{author}{L.~Q. Tran}, \bibinfo{author}{H.~T. Le},
  \bibinfo{author}{H.~V.-A. Tong},
\newblock \bibinfo{title}{Generating {{Multi-platform Single Page
  Applications}}: {{A Hierarchical Domain-Driven Design Approach}}},
\newblock in: \bibinfo{booktitle}{The 11th {{International Symposium}} on
  {{Information}} and {{Communication Technology}}}, \bibinfo{publisher}{ACM},
  \bibinfo{address}{Hanoi Vietnam}, \bibinfo{year}{2022}, pp.
  \bibinfo{pages}{344--351}. \DOIprefix\doi{10.1145/3568562.3568566}.
\bibitem[{Uluda{\u g} et~al.(2018)Uluda{\u g}, Hauder, Kleehaus, Schimpfle, and
  Matthes}]{garbajosa_supporting_2018}
\bibinfo{author}{{\"O}.~Uluda{\u g}}, \bibinfo{author}{M.~Hauder},
  \bibinfo{author}{M.~Kleehaus}, \bibinfo{author}{C.~Schimpfle},
  \bibinfo{author}{F.~Matthes},
\newblock \bibinfo{title}{Supporting {{Large-Scale Agile Development}} with
  {{Domain-Driven Design}}},
\newblock in: \bibinfo{editor}{J.~Garbajosa}, \bibinfo{editor}{X.~Wang},
  \bibinfo{editor}{A.~Aguiar} (Eds.), \bibinfo{booktitle}{Agile {{Processes}}
  in {{Software Engineering}} and {{Extreme Programming}}}, volume
  \bibinfo{volume}{314}, \bibinfo{publisher}{Springer International
  Publishing}, \bibinfo{address}{Cham}, \bibinfo{year}{2018}, pp.
  \bibinfo{pages}{232--247}. \DOIprefix\doi{10.1007/978-3-319-91602-6_16}.
\bibitem[{Wang et~al.(2022)Wang, Li, Liu, Zhang, and Pan}]{wang_reference_2022}
\bibinfo{author}{Y.~Wang}, \bibinfo{author}{S.~Li}, \bibinfo{author}{H.~Liu},
  \bibinfo{author}{H.~Zhang}, \bibinfo{author}{B.~Pan},
\newblock \bibinfo{title}{A {{Reference Architecture}} for {{Blockchain-based
  Traceability Systems Using Domain-Driven Design}} and {{Microservices}}},
\newblock in: \bibinfo{booktitle}{2022 29th {{Asia-Pacific Software Engineering
  Conference}} ({{APSEC}})}, \bibinfo{publisher}{IEEE},
  \bibinfo{address}{Japan}, \bibinfo{year}{2022}, pp.
  \bibinfo{pages}{269--278}. \DOIprefix\doi{10.1109/APSEC57359.2022.00039}.
\bibitem[{Landre et~al.(2006)Landre, Wesenberg, and
  R{\o}nneberg}]{landre_architectural_2006}
\bibinfo{author}{E.~Landre}, \bibinfo{author}{H.~Wesenberg},
  \bibinfo{author}{H.~R{\o}nneberg},
\newblock \bibinfo{title}{Architectural improvement by use of strategic level
  domain-driven design},
\newblock in: \bibinfo{booktitle}{Companion to the 21st {{ACM SIGPLAN}}
  Symposium on {{Object-oriented}} Programming Systems, Languages, and
  Applications}, \bibinfo{publisher}{ACM}, \bibinfo{address}{Portland Oregon
  USA}, \bibinfo{year}{2006}, pp. \bibinfo{pages}{809--814}.
  \DOIprefix\doi{10.1145/1176617.1176728}.
\bibitem[{Kapferer and Zimmermann(2020)}]{kapferer_domain-driven_2020}
\bibinfo{author}{S.~Kapferer}, \bibinfo{author}{O.~Zimmermann},
\newblock \bibinfo{title}{Domain-driven service design},
\newblock in: \bibinfo{booktitle}{Service-{{Oriented Computing}}},
  Communications in Computer and Information Science,
  \bibinfo{publisher}{Springer International Publishing},
  \bibinfo{address}{Cham}, \bibinfo{year}{2020}, pp. \bibinfo{pages}{189--208}.
\bibitem[{Camilli et~al.(2023)Camilli, Colarusso, Russo, and
  Zimeo}]{camilli_actor-driven_2023}
\bibinfo{author}{M.~Camilli}, \bibinfo{author}{C.~Colarusso},
  \bibinfo{author}{B.~Russo}, \bibinfo{author}{E.~Zimeo},
\newblock \bibinfo{title}{Actor-driven {{Decomposition}} of {{Microservices}}
  through {{Multi-level Scalability Assessment}}},
\newblock \bibinfo{journal}{ACM Transactions on Software Engineering and
  Methodology}  (\bibinfo{year}{2023}) \bibinfo{pages}{3583563}.
  \DOIprefix\doi{10.1145/3583563}.
\bibitem[{Braun et~al.(2021)Braun, De{\ss}loch, Wolff, Elberzhager, and
  Jedlitschka}]{braun_tackling_2021}
\bibinfo{author}{S.~Braun}, \bibinfo{author}{S.~De{\ss}loch},
  \bibinfo{author}{E.~Wolff}, \bibinfo{author}{F.~Elberzhager},
  \bibinfo{author}{A.~Jedlitschka},
\newblock \bibinfo{title}{Tackling {{Consistency-related Design Challenges}} of
  {{Distributed Data-Intensive Systems}}: {{An Action Research Study}}},
\newblock in: \bibinfo{booktitle}{Proceedings of the 15th {{ACM}} / {{IEEE
  International Symposium}} on {{Empirical Software Engineering}} and
  {{Measurement}} ({{ESEM}})}, \bibinfo{publisher}{ACM}, \bibinfo{address}{Bari
  Italy}, \bibinfo{year}{2021}, pp. \bibinfo{pages}{1--11}.
  \DOIprefix\doi{10.1145/3475716.3475771}.
\bibitem[{Landre et~al.(2007)Landre, Wesenberg, and
  Olmheim}]{landre_agile_2007}
\bibinfo{author}{E.~Landre}, \bibinfo{author}{H.~Wesenberg},
  \bibinfo{author}{J.~Olmheim},
\newblock \bibinfo{title}{Agile enterprise software development using
  domain-driven design and test first},
\newblock in: \bibinfo{booktitle}{Companion to the 22nd {{ACM SIGPLAN}}
  Conference on {{Object-oriented}} Programming Systems and Applications
  Companion}, \bibinfo{publisher}{ACM}, \bibinfo{address}{Montreal Quebec
  Canada}, \bibinfo{year}{2007}, pp. \bibinfo{pages}{983--993}.
  \DOIprefix\doi{10.1145/1297846.1297967}.
\bibitem[{Braun et~al.(2021)Braun, Bieniusa, and
  Elberzhager}]{braun_advanced_2021}
\bibinfo{author}{S.~Braun}, \bibinfo{author}{A.~Bieniusa},
  \bibinfo{author}{F.~Elberzhager},
\newblock \bibinfo{title}{Advanced {{Domain-Driven Design}} for {{Consistency}}
  in {{Distributed Data-Intensive Systems}}},
\newblock in: \bibinfo{booktitle}{Proceedings of the 8th {{Workshop}} on
  {{Principles}} and {{Practice}} of {{Consistency}} for {{Distributed Data}}},
  \bibinfo{publisher}{ACM}, \bibinfo{address}{Online United Kingdom},
  \bibinfo{year}{2021}, pp. \bibinfo{pages}{1--12}.
  \DOIprefix\doi{10.1145/3447865.3457969}.
\bibitem[{Zhao and Zhao(2021)}]{zhao_applying_2021}
\bibinfo{author}{J.~Zhao}, \bibinfo{author}{K.~Zhao},
\newblock \bibinfo{title}{Applying {{Microservice Refactoring}} to
  {{Object-2riented Legacy System}}},
\newblock in: \bibinfo{booktitle}{2021 8th {{International Conference}} on
  {{Dependable Systems}} and {{Their Applications}} ({{DSA}})},
  \bibinfo{publisher}{IEEE}, \bibinfo{address}{Yinchuan, China},
  \bibinfo{year}{2021}, pp. \bibinfo{pages}{467--473}.
  \DOIprefix\doi{10.1109/DSA52907.2021.00070}.
\bibitem[{Krause et~al.(2020)Krause, Zirkelbach, Hasselbring, Lenga, and
  Kroger}]{krause_microservice_2020}
\bibinfo{author}{A.~Krause}, \bibinfo{author}{C.~Zirkelbach},
  \bibinfo{author}{W.~Hasselbring}, \bibinfo{author}{S.~Lenga},
  \bibinfo{author}{D.~Kroger},
\newblock \bibinfo{title}{Microservice {{Decomposition}} via {{Static}} and
  {{Dynamic Analysis}} of the {{Monolith}}},
\newblock in: \bibinfo{booktitle}{2020 {{IEEE International Conference}} on
  {{Software Architecture Companion}} ({{ICSA-C}})}, \bibinfo{publisher}{IEEE},
  \bibinfo{address}{Salvador, Brazil}, \bibinfo{year}{2020}, pp.
  \bibinfo{pages}{9--16}. \DOIprefix\doi{10.1109/ICSA-C50368.2020.00011}.
\bibitem[{Snoeck and Wautelet(2022)}]{snoeck_agile_2022}
\bibinfo{author}{M.~Snoeck}, \bibinfo{author}{Y.~Wautelet},
\newblock \bibinfo{title}{Agile {{MERODE}}: A model-driven software engineering
  method for user-centric and value-based development},
\newblock \bibinfo{journal}{Software and Systems Modeling} \bibinfo{volume}{21}
  (\bibinfo{year}{2022}) \bibinfo{pages}{1469--1494}.
  \DOIprefix\doi{10.1007/s10270-022-01015-y}.
\bibitem[{Danenas and Garsva(2012)}]{skersys_domain_2012}
\bibinfo{author}{P.~Danenas}, \bibinfo{author}{G.~Garsva},
\newblock \bibinfo{title}{Domain {{Driven Development}} and {{Feature Driven
  Development}} for {{Development}} of {{Decision Support Systems}}},
\newblock in: \bibinfo{editor}{T.~Skersys}, \bibinfo{editor}{R.~Butleris},
  \bibinfo{editor}{R.~Butkiene} (Eds.), \bibinfo{booktitle}{Information and
  {{Software Technologies}}}, volume \bibinfo{volume}{319},
  \bibinfo{publisher}{Springer Berlin Heidelberg}, \bibinfo{address}{Berlin,
  Heidelberg}, \bibinfo{year}{2012}, pp. \bibinfo{pages}{187--198}.
  \DOIprefix\doi{10.1007/978-3-642-33308-8_16}.
\bibitem[{Kapferer and Zimmermann(2021)}]{hammoudi_domain-driven_2021}
\bibinfo{author}{S.~Kapferer}, \bibinfo{author}{O.~Zimmermann},
\newblock \bibinfo{title}{Domain-{{Driven Architecture Modeling}} and {{Rapid
  Prototyping}} with {{Context Mapper}}},
\newblock in: \bibinfo{editor}{S.~Hammoudi}, \bibinfo{editor}{L.~F. Pires},
  \bibinfo{editor}{B.~Seli{\'c}} (Eds.), \bibinfo{booktitle}{Model-{{Driven
  Engineering}} and {{Software Development}}}, volume \bibinfo{volume}{1361},
  \bibinfo{publisher}{Springer International Publishing},
  \bibinfo{address}{Cham}, \bibinfo{year}{2021}, pp. \bibinfo{pages}{250--272}.
  \DOIprefix\doi{10.1007/978-3-030-67445-8_11}.
\bibitem[{Hammarstr{\"o}m and Herzog(2016)}]{hammarstrom_experience_2016}
\bibinfo{author}{P.~Hammarstr{\"o}m}, \bibinfo{author}{E.~Herzog},
\newblock \bibinfo{title}{Experience from integrating {{Domain Driven Software
  System Design}} into a {{Systems Engineering Organization}}},
\newblock \bibinfo{journal}{INCOSE International Symposium}
  \bibinfo{volume}{26} (\bibinfo{year}{2016}) \bibinfo{pages}{1192--1203}.
  \DOIprefix\doi{10.1002/j.2334-5837.2016.00220.x}.
\bibitem[{Da~Silva et~al.(2022)Da~Silva, Gomes, and
  Basu}]{da_silva_bpm2ddd_2022}
\bibinfo{author}{C.~E. Da~Silva}, \bibinfo{author}{E.~L. Gomes},
  \bibinfo{author}{S.~S. Basu},
\newblock \bibinfo{title}{{{BPM2DDD}}: {{A Systematic Process}} for
  {{Identifying Domains}} from {{Business Processes Models}}},
\newblock \bibinfo{journal}{Software} \bibinfo{volume}{1}
  (\bibinfo{year}{2022}) \bibinfo{pages}{417--449}.
  \DOIprefix\doi{10.3390/software1040018}.
\bibitem[{Hippchen et~al.(2017)Hippchen, Giessler, Steinegger, Schneider, and
  Abeck}]{hippchen_designing_2017}
\bibinfo{author}{B.~Hippchen}, \bibinfo{author}{P.~Giessler},
  \bibinfo{author}{R.~H. Steinegger}, \bibinfo{author}{M.~Schneider},
  \bibinfo{author}{S.~Abeck},
\newblock \bibinfo{title}{Designing {{Microservice-Based Applications}} by
  {{Using}} a {{Domain-Driven Design Approach}}}  (\bibinfo{year}{2017}).
\bibitem[{Ding et~al.(2020)Ding, Wang, Li, Wang, and
  Zhang}]{ding_enterprise_2020}
\bibinfo{author}{Y.~Ding}, \bibinfo{author}{L.~Wang}, \bibinfo{author}{S.~Li},
  \bibinfo{author}{X.~Wang}, \bibinfo{author}{J.~Zhang},
\newblock \bibinfo{title}{Enterprise service application architecture based on
  {{Domain Driven Model Design}}},
\newblock in: \bibinfo{booktitle}{2020 2nd {{International Conference}} on
  {{Information Technology}} and {{Computer Application}} ({{ITCA}})},
  \bibinfo{publisher}{IEEE}, \bibinfo{address}{Guangzhou, China},
  \bibinfo{year}{2020}, pp. \bibinfo{pages}{778--784}.
  \DOIprefix\doi{10.1109/ITCA52113.2020.00167}.
\bibitem[{{\"O}zkan et~al.(2023){\"O}zkan, Babur, and Van
  Den~Brand}]{ozkan_refactoring_2023}
\bibinfo{author}{O.~{\"O}zkan}, \bibinfo{author}{{\"O}.~Babur},
  \bibinfo{author}{M.~Van Den~Brand},
\newblock \bibinfo{title}{Refactoring with domain-driven design in an
  industrial context: {{An}} action research report},
\newblock \bibinfo{journal}{Empirical Software Engineering}
  \bibinfo{volume}{28} (\bibinfo{year}{2023}) \bibinfo{pages}{94}.
  \DOIprefix\doi{10.1007/s10664-023-10310-1}.
\bibitem[{Pereira and Silva(2022)}]{pereira_towards_2022}
\bibinfo{author}{P.~Pereira}, \bibinfo{author}{A.~R. Silva},
  \bibinfo{title}{Towards {{Transactional Causal Consistent Microservices
  Business Logic}}}, \bibinfo{year}{2022}.
  \href{http://arxiv.org/abs/2212.11658}{{\tt arXiv:2212.11658}}.
\bibitem[{Joselyne et~al.(2021)Joselyne, Bajpai, and
  Nzanywayingoma}]{joselyne_systematic_2021}
\bibinfo{author}{M.~I. Joselyne}, \bibinfo{author}{G.~Bajpai},
  \bibinfo{author}{F.~Nzanywayingoma},
\newblock \bibinfo{title}{A {{Systematic Framework}} of {{Application
  Modernization}} to {{Microservice}} based {{Architecture}}},
\newblock in: \bibinfo{booktitle}{2021 {{International Conference}} on
  {{Engineering}} and {{Emerging Technologies}} ({{ICEET}})},
  \bibinfo{publisher}{IEEE}, \bibinfo{address}{Istanbul, Turkey},
  \bibinfo{year}{2021}, pp. \bibinfo{pages}{1--6}.
  \DOIprefix\doi{10.1109/ICEET53442.2021.9659783}.
\bibitem[{Wanderley and Da~Silveria(2012)}]{wanderley_framework_2012}
\bibinfo{author}{F.~Wanderley}, \bibinfo{author}{D.~S. Da~Silveria},
\newblock \bibinfo{title}{A {{Framework}} to {{Diminish}} the {{Gap}} between
  the {{Business Specialist}} and the {{Software Designer}}},
\newblock in: \bibinfo{booktitle}{2012 {{Eighth International Conference}} on
  the {{Quality}} of {{Information}} and {{Communications Technology}}},
  \bibinfo{publisher}{IEEE}, \bibinfo{address}{Lisbon, TBD, Portugal},
  \bibinfo{year}{2012}, pp. \bibinfo{pages}{199--204}.
  \DOIprefix\doi{10.1109/QUATIC.2012.9}.
\bibitem[{Koryl(2017)}]{koryl_active_2017}
\bibinfo{author}{M.~Koryl},
\newblock \bibinfo{title}{Active resources concept of computation for
  enterprise software},
\newblock \bibinfo{journal}{Archives of Control Sciences} \bibinfo{volume}{27}
  (\bibinfo{year}{2017}) \bibinfo{pages}{279--291}.
  \DOIprefix\doi{10.1515/acsc-2017-0018}.
\bibitem[{Mayer and Weinreich(2018)}]{mayer_approach_2018}
\bibinfo{author}{B.~Mayer}, \bibinfo{author}{R.~Weinreich},
\newblock \bibinfo{title}{An {{Approach}} to {{Extract}} the {{Architecture}}
  of {{Microservice-Based Software Systems}}},
\newblock in: \bibinfo{booktitle}{2018 {{IEEE Symposium}} on {{Service-Oriented
  System Engineering}} ({{SOSE}})}, \bibinfo{publisher}{IEEE},
  \bibinfo{address}{Bamberg}, \bibinfo{year}{2018}, pp.
  \bibinfo{pages}{21--30}. \DOIprefix\doi{10.1109/SOSE.2018.00012}.
\bibitem[{Vural and Koyuncu(2021)}]{vural_does_2021}
\bibinfo{author}{H.~Vural}, \bibinfo{author}{M.~Koyuncu},
\newblock \bibinfo{title}{Does {{Domain-Driven Design Lead}} to {{Finding}} the
  {{Optimal Modularity}} of a {{Microservice}}?},
\newblock \bibinfo{journal}{IEEE Access} \bibinfo{volume}{9}
  (\bibinfo{year}{2021}) \bibinfo{pages}{32721--32733}.
  \DOIprefix\doi{10.1109/ACCESS.2021.3060895}.
\bibitem[{Oukes et~al.(2021)Oukes, Andel, Folmer, Bennett, and
  Lemmen}]{oukes_domain-driven_2021}
\bibinfo{author}{P.~Oukes}, \bibinfo{author}{M.~V. Andel},
  \bibinfo{author}{E.~Folmer}, \bibinfo{author}{R.~Bennett},
  \bibinfo{author}{C.~Lemmen},
\newblock \bibinfo{title}{Domain-{{Driven Design}} applied to land
  administration system development: {{Lessons}} from the {{Netherlands}}},
\newblock \bibinfo{journal}{Land Use Policy} \bibinfo{volume}{104}
  (\bibinfo{year}{2021}) \bibinfo{pages}{105379}.
  \DOIprefix\doi{10.1016/j.landusepol.2021.105379}.
\bibitem[{H{\"a}ser et~al.(2016)H{\"a}ser, Felderer, and Breu}]{haser_is_2016}
\bibinfo{author}{F.~H{\"a}ser}, \bibinfo{author}{M.~Felderer},
  \bibinfo{author}{R.~Breu},
\newblock \bibinfo{title}{Is business domain language support beneficial for
  creating test case specifications: {{A}} controlled experiment},
\newblock \bibinfo{journal}{Information and Software Technology}
  \bibinfo{volume}{79} (\bibinfo{year}{2016}) \bibinfo{pages}{52--62}.
  \DOIprefix\doi{10.1016/j.infsof.2016.07.001}.
\bibitem[{Le et~al.(2018)Le, Dang, and Nguyen}]{le_domain_2018}
\bibinfo{author}{D.~M. Le}, \bibinfo{author}{D.-H. Dang},
  \bibinfo{author}{V.-H. Nguyen},
\newblock \bibinfo{title}{On domain driven design using annotation-based domain
  specific language},
\newblock \bibinfo{journal}{Computer Languages, Systems \& Structures}
  \bibinfo{volume}{54} (\bibinfo{year}{2018}) \bibinfo{pages}{199--235}.
  \DOIprefix\doi{10.1016/j.cl.2018.05.001}.
\bibitem[{Abeck et~al.(2019)Abeck, Schneider, Quirmbach, Klarl, Urbaczek, and
  Zogaj}]{abeck_context_2019}
\bibinfo{author}{S.~Abeck}, \bibinfo{author}{M.~Schneider},
  \bibinfo{author}{J.-P. Quirmbach}, \bibinfo{author}{H.~Klarl},
  \bibinfo{author}{C.~Urbaczek}, \bibinfo{author}{S.~Zogaj},
\newblock \bibinfo{title}{A {{Context Map}} as the {{Basis}} for a
  {{Microservice Architecture}} for the {{Connected Car Domain}}}
  (\bibinfo{year}{2019}). \DOIprefix\doi{10.18420/INF2019_18}.
\bibitem[{B{\"u}nder and Kuchen(2019)}]{bunder_model-driven_2019}
\bibinfo{author}{H.~B{\"u}nder}, \bibinfo{author}{H.~Kuchen},
\newblock \bibinfo{title}{A model-driven approach for behavior-driven {{GUI}}
  testing},
\newblock in: \bibinfo{booktitle}{Proceedings of the 34th {{ACM}}/{{SIGAPP
  Symposium}} on {{Applied Computing}}}, \bibinfo{publisher}{ACM},
  \bibinfo{address}{Limassol Cyprus}, \bibinfo{year}{2019}, pp.
  \bibinfo{pages}{1742--1751}. \DOIprefix\doi{10.1145/3297280.3297450}.
\bibitem[{Dragoni et~al.(2017)Dragoni, Giallorenzo, Lafuente, Mazzara, Montesi,
  Mustafin, and Safina}]{dragoni_microservices_2017}
\bibinfo{author}{N.~Dragoni}, \bibinfo{author}{S.~Giallorenzo},
  \bibinfo{author}{A.~L. Lafuente}, \bibinfo{author}{M.~Mazzara},
  \bibinfo{author}{F.~Montesi}, \bibinfo{author}{R.~Mustafin},
  \bibinfo{author}{L.~Safina}, \bibinfo{title}{Microservices: Yesterday, today,
  and tomorrow}, \bibinfo{year}{2017}.
  \DOIprefix\doi{10.48550/arXiv.1606.04036}.
  \href{http://arxiv.org/abs/1606.04036}{{\tt arXiv:1606.04036}}.
\bibitem[{Fritzsch et~al.(2019)Fritzsch, Bogner, Wagner, and
  Zimmermann}]{fritzsch_microservices_2019}
\bibinfo{author}{J.~Fritzsch}, \bibinfo{author}{J.~Bogner},
  \bibinfo{author}{S.~Wagner}, \bibinfo{author}{A.~Zimmermann},
\newblock \bibinfo{title}{Microservices {{Migration}} in {{Industry}}:
  {{Intentions}}, {{Strategies}}, and {{Challenges}}},
\newblock in: \bibinfo{booktitle}{2019 {{IEEE International Conference}} on
  {{Software Maintenance}} and {{Evolution}} ({{ICSME}})},
  \bibinfo{publisher}{IEEE}, \bibinfo{address}{Cleveland, OH, USA},
  \bibinfo{year}{2019}, pp. \bibinfo{pages}{481--490}.
  \DOIprefix\doi{10.1109/ICSME.2019.00081}.
\bibitem[{Lewis and Fowler(2014)}]{lewis2014microservices}
\bibinfo{author}{J.~Lewis}, \bibinfo{author}{M.~Fowler},
\newblock \bibinfo{title}{Microservices: A definition of this new architectural
  term},
\newblock \bibinfo{journal}{MartinFowler. com} \bibinfo{volume}{25}
  (\bibinfo{year}{2014}) \bibinfo{pages}{12}.
\bibitem[{Hamzehloui et~al.(2018)Hamzehloui, Sahibuddin, and
  Salah}]{hamzehloui_systematic_2018}
\bibinfo{author}{M.~S. Hamzehloui}, \bibinfo{author}{S.~Sahibuddin},
  \bibinfo{author}{K.~Salah},
\newblock \bibinfo{title}{A {{Systematic Mapping Study}} on {{Microservices}}},
\newblock in: \bibinfo{booktitle}{Advances in {{Intelligent Systems}} and
  {{Computing}}}, \bibinfo{publisher}{Springer International Publishing},
  \bibinfo{year}{2018}, pp. \bibinfo{pages}{1079--1090}.
  \DOIprefix\doi{10.1007/978-3-319-99007-1_100}.
\bibitem[{Viggiato et~al.(2018)Viggiato, Terra, Rocha, Valente, and
  Figueiredo}]{viggiato_microservices_2018}
\bibinfo{author}{M.~Viggiato}, \bibinfo{author}{R.~Terra},
  \bibinfo{author}{H.~Rocha}, \bibinfo{author}{M.~T. Valente},
  \bibinfo{author}{E.~Figueiredo},
\newblock \bibinfo{title}{Microservices in {{Practice}}: {{A Survey Study}}}
  (\bibinfo{year}{2018}). \DOIprefix\doi{10.48550/ARXIV.1808.04836}.
\bibitem[{Knoche and Hasselbring(2019)}]{knoche_drivers_2019}
\bibinfo{author}{H.~Knoche}, \bibinfo{author}{W.~Hasselbring},
\newblock \bibinfo{title}{Drivers and {{Barriers}} for {{Microservice
  Adoption}} -- {{A Survey}} among {{Professionals}} in {{Germany}}},
\newblock \bibinfo{journal}{Enterprise Modelling and Information Systems
  Architectures (EMISAJ)}  (\bibinfo{year}{2019}) \bibinfo{pages}{1:1--35
  Pages}. \DOIprefix\doi{10.18417/EMISA.14.1}.
\bibitem[{Voinov and Bousquet(2010)}]{voinov_modeling_2010}
\bibinfo{author}{A.~Voinov}, \bibinfo{author}{F.~Bousquet},
\newblock \bibinfo{title}{Modeling with stakeholders},
\newblock \bibinfo{journal}{Environmental Modelling \& Software}
  \bibinfo{volume}{25} (\bibinfo{year}{2010}) \bibinfo{pages}{1268--1281}.
\bibitem[{Brandolini(2019)}]{brandolini_eventstorming_2019}
\bibinfo{author}{A.~Brandolini}, \bibinfo{title}{Introducing EventStorming: An
  act of deliberate collective learning}, \bibinfo{publisher}{Leanpub},
  \bibinfo{year}{2019}.
\bibitem[{Famelis et~al.(2012)Famelis, Salay, and
  Chechik}]{famelis_partial_2012}
\bibinfo{author}{M.~Famelis}, \bibinfo{author}{R.~Salay},
  \bibinfo{author}{M.~Chechik},
\newblock \bibinfo{title}{Partial models: Towards modeling and reasoning with
  uncertainty},
\newblock in: \bibinfo{booktitle}{Proceedings of the 34th International
  Conference on Software Engineering (ICSE)}, \bibinfo{organization}{IEEE},
  \bibinfo{year}{2012}, pp. \bibinfo{pages}{573--583}.
\bibitem[{Babur et~al.(2024)Babur, Constantinou, and
  Serebrenik}]{babur2024language}
\bibinfo{author}{{\"O}.~Babur}, \bibinfo{author}{E.~Constantinou},
  \bibinfo{author}{A.~Serebrenik},
\newblock \bibinfo{title}{Language usage analysis for emf metamodels on
  github},
\newblock \bibinfo{journal}{Empirical Software Engineering}
  \bibinfo{volume}{29} (\bibinfo{year}{2024}) \bibinfo{pages}{23}.
\bibitem[{Petersen et~al.(2015)Petersen, Vakkalanka, and
  Kuzniarz}]{petersen_guidelines_2015}
\bibinfo{author}{K.~Petersen}, \bibinfo{author}{S.~Vakkalanka},
  \bibinfo{author}{L.~Kuzniarz},
\newblock \bibinfo{title}{Guidelines for conducting systematic mapping studies
  in software engineering: {{An}} update},
\newblock \bibinfo{journal}{Information and Software Technology}
  \bibinfo{volume}{64} (\bibinfo{year}{2015}) \bibinfo{pages}{1--18}.
  \DOIprefix\doi{10.1016/j.infsof.2015.03.007}.
\bibitem[{Petersen et~al.(2008)Petersen, Feldt, Mujtaba, and
  Mattsson}]{petersen_systematic_2008}
\bibinfo{author}{K.~Petersen}, \bibinfo{author}{R.~Feldt},
  \bibinfo{author}{S.~Mujtaba}, \bibinfo{author}{M.~Mattsson},
\newblock \bibinfo{title}{Systematic {{Mapping Studies}} in {{Software
  Engineering}}},
\newblock in: \bibinfo{booktitle}{12th {{International Conference}} on
  {{Evaluation}} and {{Assessment}} in {{Software Engineering}} ({{EASE}})},
  \bibinfo{year}{2008}. \DOIprefix\doi{10.14236/ewic/EASE2008.8}.

\end{thebibliography}




\end{document}